\documentclass[12pt]{article}

\usepackage[margin=1in]{geometry}
\usepackage{fontspec}
\usepackage{lmodern}
\usepackage{setspace}
\usepackage{amsmath}
\usepackage{booktabs}
\usepackage{dcolumn}
\usepackage{longtable}
\usepackage{threeparttable}
\usepackage{graphicx}
\usepackage{caption}
\usepackage{subcaption}
\usepackage{csquotes}
\usepackage[
  backend=biber,
  style=apa,
  natbib=true,
  maxcitenames=2,
  maxbibnames=99,
  uniquelist=false,
  uniquename=false,
  doi=false,
  isbn=false,
  url=false
]{biblatex}
\usepackage[colorlinks=true,linkcolor=blue,citecolor=blue,urlcolor=blue]{hyperref}

\newcommand{\sizeslope}{\hat{\beta}^{\mathrm{size}}}
\newcommand{\trust}{\mathrm{Trust}}
\newcommand{\rol}{\mathrm{RuleOfLaw}}

\title{Trust, Rule of Law, and the Size Premium:\\Evidence from a Meta-Analysis\thanks{Corresponding author: Jiri Schwarz, Anglo-American University, Prague; j.schwarz.sbu@aauni.edu. The full replication package (data, code, and the online appendix) is available at \url{https://meta-analysis.cz/trust} and archived at \url{https://doi.org/10.5281/zenodo.21486177}. The authors acknowledge support from the Czech Science Foundation (grant 24-11583S).}}

\author{%
Jiri Schwarz\textsuperscript{a}, Tomas Havranek\textsuperscript{b,c,d}, Zuzana Irsova\textsuperscript{a}, and Jiri Novak\textsuperscript{b}\\[0.5em]
{\small\textsuperscript{a}Anglo-American University, Prague}\\[-0.4em]
{\small\textsuperscript{b}Institute of Economic Studies, Charles University, Prague}\\[-0.4em]
{\small\textsuperscript{c}Centre for Economic Policy Research, London}\\[-0.4em]
{\small\textsuperscript{d}Meta-Research Innovation Center, Stanford}
}
\date{August 9, 2026}

\begin{document}

\maketitle

\vspace{-2em}

{\singlespacing\small
\begin{abstract}
\noindent Reported estimates of the size premium, the tendency of smaller firms to earn higher average returns than larger firms, vary widely across studies, countries, periods, and designs. We examine whether generalized trust and rule of law help account for that heterogeneity. Small firms are more opaque and more dependent on outside finance, so the enforcement and information environment should matter more for them than for large firms. We study 1,613 reported size-slope estimates from 105 studies and 31 countries. The meta-regressions control for study design, specification, precision, publication context, and market and macro-financial conditions; Bayesian model averaging assesses uncertainty over the control set. The more stable association is with rule of law, and it runs against the intuitive expectation that better legal institutions shrink the premium: stronger rule of law is associated with more negative reported size slopes, hence larger conventional size premia. The trust association is conditional and less precisely estimated: where rule of law is weak, higher generalized trust is linked to less negative reported slopes (and thus a weaker premium), and this link fades as rule of law strengthens. Formal and informal institutions thus help organize part of the disagreement in this literature, although the analysis concerns variation in reported estimates and does not identify causal effects.
\end{abstract}

\noindent\textbf{Keywords:} size premium; trust; rule of law; meta-analysis; Bayesian model averaging

\medskip

\noindent\textbf{JEL codes:} G12; G15; C83
}

\clearpage

\section{Introduction}
The size premium is one of the oldest empirical regularities in asset pricing \citep{banz1981,fama1992}, yet its reported magnitude differs sharply across markets, periods, and research designs. We ask whether generalized trust and rule of law help organize the disagreement in what the literature reports.

Since the early evidence on size, many studies have reported that small firms earn higher average returns than large firms. Later work has questioned the stability, interpretation, and independent role of size once measurement choices, market microstructure, other anomalies, and data-mining concerns are taken seriously \citep{roll1981,schwert2003,hou2020}. A researcher looking for a single calibration number, or for a simple verdict on the anomaly, quickly encounters a literature in which plausible estimates point in different directions.

Institutions are a candidate source of that heterogeneity because small firms are especially exposed to frictions that formal rules and informal norms can amplify or soften. Smaller firms tend to be more opaque, less collateralized, more dependent on local finance, and less able to absorb the fixed costs of compliance, disclosure, litigation, and contract enforcement \citep{berger1998,beck2005b,beck2008,hadlock2010,iliev2010}. Rule of law captures the formal side of this environment: predictable courts, secure property rights, investor protection, and credible enforcement of contracts, which can reduce uncertainty, deepen markets, and support outside finance \citep{north1990,laporta1998,laporta2006,acemoglu2005,hail2006,laporta2002,stulz1999,stulz2005,levine2005}. But its effect on the size premium is not one-directional: stronger legal and regulatory systems may also make small-firm risks more visible, impose fixed compliance burdens that fall more heavily on smaller firms, and price small-firm risk more cleanly \citep{leuz2007,gao2009,iliev2010}.

Generalized trust captures a different, informal side of the same environment. Informal constraints and social norms shape exchange alongside formal law \citep{north1990,greif1994,tabellini2010}. Trust can lower the cost of transacting when contracts are incomplete, monitoring is costly, or counterparties are hard to verify \citep{arrow1972,coleman1988,putnam1993,fukuyama1995}, even if some of what surveys call trust reflects calculated expectations about risk \citep{williamson1993}. For financial markets the empirical link is well documented: generalized trust is associated with stock market participation, credit access, disclosure credibility, and financing terms \citep{guiso2004,guiso2008,guiso2009,pevzner2015,bottazzi2016}. These channels should matter most for small and opaque firms, whose investors and lenders rely more heavily on soft information, perceived honesty, and informal assurance \citep{berger1995,beck2005b,hasan2017,gupta2018}. This is why the analysis focuses on generalized trust and treats family trust separately as a placebo measure of particularized, non-market trust \citep{banfield1958,uslaner2002,delhey2011,alesina2010}.

Cross-country asset-pricing studies document that the size effect differs across markets \citep{fama2012,fama2017,mishra2022}, and institutional finance links trust and legal quality to participation, disclosure, and the cost of capital. Yet these literatures have not, to our knowledge, tested whether generalized trust and rule of law account for part of the heterogeneity in reported size-premium estimates. We assemble an estimate-level dataset of 1,613 reported size slopes and match each estimate to generalized trust from the European Values Study and World Values Survey (EVS/WVS) and rule of law from the Worldwide Governance Indicators (WGI). The size coefficients are drawn from the inventory coded by \citet{astakhov2019}; the institutional layer and the analysis are new. We estimate the direct trust and rule-of-law associations in reported size slopes. We then test whether the trust association changes with formal legal quality, conditioning on study design, specification choices, estimate precision, and macro-financial conditions.

The substitution logic yields a conditional prediction: generalized trust may substitute for weak formal enforcement by making investors and creditors more willing to transact with small, opaque firms even when legal protection is limited \citep{guiso2004,aghion2010,pevzner2015,li2019}. In such environments, higher trust should be associated with a smaller size premium. As rule of law improves, the marginal role of trust should fade because formal institutions already provide more of the enforcement and disclosure infrastructure that market participants need \citep{ahlerup2009,bjrnskov2022}. The proposed substitution is specific to the small-firm financing margin and yields no causal claim about the true size premium. Other settings find complementarity or mixed interactions between trust and law \citep{bloom2012,carlin2009,cruzgarcia2019,bartling2025}. Our empirical prediction is correspondingly narrow: generalized trust should have its strongest association with reported size coefficients where rule of law is weaker.

The intuitive expectation is that better legal institutions ease the frictions that bind small firms and so shrink the premium. The data say the opposite. The clearest result is that stronger rule of law is associated with more negative reported size slopes, that is, with larger conventional size premia, and the association holds even after excluding estimates from the United States. It is consistent with cleaner pricing of small-firm risk and with fixed compliance costs that fall regressively on size. Generalized trust adds a second, more tentative pattern: it is associated with less negative slopes where rule of law is weak, and this association fades as rule of law improves. The interaction is imprecise under clustered inference and thins out without the U.S. sample. Bayesian model averaging preserves the same posterior signs under control-set uncertainty. These are associations in reported estimates, not causal effects. Section 2 sets out the mechanism; Sections 3--6 take it to the data; Sections 7 and 8 interpret and conclude.

The data and code used to reproduce every table and figure are available in an online appendix at \url{https://meta-analysis.cz/trust} and archived at \url{https://doi.org/10.5281/zenodo.21486177}. The package starts from the frozen, analysis-ready estimate-level data, runs with a single command, and reproduces every reported number, so the analysis can be inspected from the matched-data stage onward. The paper follows the MAER-Net reporting guidelines for meta-analysis in economics as updated for artificial intelligence \citep{cook2026reporting}, together with the accompanying principles for AI use \citep{cook2026ai}. Appendix \ref{app:reporting} records where the requirements are met and lists the points on which we depart from them, most of which follow from starting at the estimate level rather than from a new literature search.

\section{Background and Hypotheses}\label{sec:background}

\subsection{The Size Premium and Reported Estimates}
The size premium refers to the tendency for smaller firms to earn higher average returns than larger firms. What the literature disputes is the magnitude and stability of that gap across markets, periods, and specifications. The estimates used in this analysis are reported slopes on firm size, so more negative reported size coefficients imply a stronger size premium throughout this paper. \citet{banz1981} first documented a negative relation between firm size and stock returns that could not be explained by market beta alone \citep{sharpe1964,lintner1965,black1972}. \citet{reinganum1981} and \citet{keim1983} showed that the pattern also appeared alongside other cross-sectional anomalies and displayed distinctive seasonal and small-firm concentration features. \citet{fama1992,fama1993} later placed the effect at the center of the modern factor-pricing literature through the small-minus-big factor, and size remains one of the canonical characteristics used to describe the cross-section \citep{harvey2016}. That canonical status matters here because it establishes firm size as a recurring dimension of priced return differences rather than as a one-off anomaly confined to a single market or period.

The literature also portrays the size premium as unstable across periods, markets, and empirical specifications. Debate continues over whether the premium is weak, hidden, affected by measurement design, or absorbed by other characteristics \citep{roll1981,schwert2003}. \citet{vandijk2011} reviews the large body of evidence and emphasizes its uneven strength across countries. That diversity appears in both developed and emerging markets: \citet{barry2002} find that size and value effects are less robust in emerging equity markets than in the canonical U.S. evidence, while \citet{hou2011} show that the set of empirically successful return predictors differs materially across countries. \citet{fama2012,fama2017} likewise show that the role of size varies across international markets and factor-model specifications. Even within the U.S. literature, the time profile is disputed: some evidence points to attenuation after the original discovery, whereas other work argues that expected and realized size premia can diverge because small firms are hit by unfavorable profitability shocks \citep{asness2018,hou2019}. The unresolved debate leaves substantial heterogeneity in measured size effects as the central fact to be explained; the institutional environment is one candidate explanation for it.

The reported estimates themselves have been collected at the estimate level by \citet{astakhov2019}, who document publication bias, study-design effects, and substantial residual heterogeneity. Two facts from that inventory matter here: the literature is large enough to support a formal meta-regression, and the variation in reported size effects is a real empirical pattern that invites systematic explanation. We do not re-estimate publication bias; the residual heterogeneity is what we try to explain.

Our object of analysis is the set of reported estimate-level size coefficients. Their variation may reflect economic heterogeneity as well as differences in sample construction, econometric specification, publication incentives, and other features of the research process. We ask whether part of the remaining variation lines up with the institutional environments in which small firms raise capital, disclose information, and bear enforcement frictions. Generalized trust and rule of law capture two relevant features of those environments.

\subsection{Trust as an Informal Institution}

If the heterogeneity in reported size coefficients partly reflects the institutional environments in which firms seek finance, then generalized trust is a natural place to begin. In the institutional-economics tradition, trust belongs to the domain of informal institutions: it is embedded in social norms, expectations, and repeated patterns of behavior that shape how readily economic actors engage in impersonal exchange \citep{north1990}. This is why the classical trust literature treats it as economically consequential rather than merely sociological background. \citet{arrow1972} argues that trust is present in virtually every commercial transaction because contracts are incomplete and opportunism is costly to police ex ante. \citet{coleman1988,coleman1990} frames trust as a form of social capital that lowers monitoring and agency costs, while \citet{putnam1993,putnam2000} and \citet{fukuyama1995} connect generalized trust to the broader capacity of societies to sustain cooperation beyond narrow personal networks. When exchange depends on the willingness of strangers to transact under imperfect information, trust can plausibly affect financing conditions and, through them, reported size coefficients.

That claim depends on distinguishing generalized trust from particularized trust. For financial markets the relevant concept is the broader belief that unknown counterparties will usually honor obligations and refrain from exploitation, rather than trust confined to family, kin, or familiar in-groups. This distinction runs from \citeauthor{banfield1958}'s \parencite*{banfield1958} contrast between civic cooperation and ``amoral familism'' to \citeauthor{uslaner2002}'s \parencite*{uslaner2002} emphasis on moralistic trust. It continues in the radius-of-trust concept of \citet{delhey2011} and in \citeauthor{alesina2010}'s \parencite*{alesina2010} evidence that strong family ties can substitute for, rather than reinforce, generalized trust. It is also consistent with the organizational evidence in \citet{laporta1997trust} and the historical contrast in \citet{greif1994} between trust confined to narrow groups and institutions that support broader anonymous exchange. This distinction is central here because asset markets are arm's-length settings. They require investors, lenders, and intermediaries to deal with firms and managers they do not know personally.

The concept nevertheless requires discipline. \citet{williamson1993} warns that some of what researchers label trust may instead be calculative expectations about the probability of defection. That caution is useful, but it does not eliminate the empirical relevance of survey-based generalized trust. Cross-country evidence shows that such measures predict economic performance and development over and above simple proxies for risk-taking or contemporary formal institutions \citep{knack1997,algan2010,tabellini2010}. In finance, the evidence is more specific. \citet{guiso2004} show that social capital is associated with greater use of formal financial instruments and institutional credit, \citet{guiso2008} trace part of the stock-market participation margin to the perceived probability of being cheated, and \citet{guiso2009} extend the same logic to international exchange. Related work links trust to greater household participation in equity markets, more credible disclosure, and less frictive financing relationships \citep{georgarakos2011,bricker2023,pevzner2015,bottazzi2016}. These studies point to an economic role for generalized trust: it plausibly lowers the cost of initiating and monitoring financial transactions between parties who do not know one another.

Crisis-period corporate finance points the same way: firms with stronger CSR capital earned higher stock returns in the 2008--2009 crisis \citep{lins2017} and could issue bonds at lower spreads and longer maturities \citep{amiraslani2023}. CSR is not identical to generalized social trust, but both patterns are consistent with trust-relevant reputational capital lowering financing premia when formal contracting and verification are under stress.

That mechanism should matter especially for small firms. The previous subsection emphasized that reported size coefficients are likely to vary with the frictions faced by firms that are more opaque, less collateralized, and more dependent on local or relationship-based finance. Trust fits naturally into that margin because it is most valuable when hard information is scarce and counterparties must rely more heavily on soft information, reputation, and perceived honesty. The most direct supporting evidence comes from \citet{hasan2017}, who show that higher local social capital is associated with lower loan spreads and bond yields. \citet{gupta2018} similarly link social capital to a lower implied cost of equity, and \citet{li2019} show that trust reduces IPO underpricing particularly for smaller and growth-oriented firms. These are not direct tests of the size premium itself, but they motivate our test of whether trust covaries with reported size slopes. If trust disproportionately relaxes the financing and information frictions borne by small firms, variation in generalized trust is a credible candidate for explaining heterogeneity in reported size-premium estimates.

\subsection{Rule of Law as a Formal Institution}

If trust captures the informal side of market exchange, rule of law captures the formal side. In institutional economics, formal rules matter because impersonal exchange does not scale without credible enforcement of contracts, predictable adjudication, and reasonably secure property rights \citep{north1990,acemoglu2001,acemoglu2005b}. The law-and-finance literature translates that general insight into a financial-markets setting. \citet{laporta1997finance,laporta1998,laporta1999,laporta2000,laporta2002} show that legal origin, investor protection, and enforcement quality are closely linked to ownership structures, market development, valuation, and the ability of outside investors to supply capital. Later work emphasizes that disclosure obligations and enforcement standards are core parts of the institutional environment that make securities markets usable for dispersed investors \citep{laporta2006,laporta2008,djankov2003,djankov2008}. That is why we use rule of law, rather than a broader governance measure, as the formal-institution anchor. It maps most directly to the set of formal arrangements that should matter for reported size coefficients: contract enforcement, shareholder protection, legal recourse, and the credibility of disclosure.

These mechanisms suggest an obvious channel through which stronger rule of law could reduce the size premium. Small firms are typically more opaque, more locally exposed, and more dependent on external finance that is difficult to secure when investors fear expropriation or weak enforcement. Improvements in formal legal institutions can therefore relax frictions that bind more tightly for smaller firms than for larger ones. Cross-country studies show that better legal environments and financial development expand firms' access to long-term external finance, lower the sensitivity of investment to internal funds, and reduce the financing obstacles reported by the smallest firms in particular \citep{demirguckunt1998,demirguckunt2002,beck2003,beck2005,beck2005b,beck2008,love2003}. The same logic appears in asset-pricing and governance evidence: stronger legal institutions are associated with lower costs of equity, more credible access to outside capital, and less severe governance penalties for firms that cannot easily contract around domestic institutional weakness \citep{hail2006,hail2009,stulz1999,stulz2005,doidge2007,himmelberg2004}. Through this channel, better rule of law should compress the size premium by reducing the financing, contracting, and information disadvantages that fall most heavily on small firms.

But that prediction is not the only one supported by the literature. Stronger legal environments can also make the reported size effect more pronounced. One reason is that the same legal architecture that protects investors also imposes fixed disclosure, auditing, and compliance costs whose incidence is regressive in firm size. \citet{iliev2010} shows that tighter internal-control compliance requirements under Sarbanes-Oxley generated substantial cost increases and negative valuation effects around the relevant threshold. \citet{gao2009} document that firms respond strategically to avoid those burdens. More generally, stronger enforcement can raise the informational precision with which small-firm risk is priced. Better disclosure regimes, lower earnings opacity, and more credible enforcement may reduce noise without eliminating the underlying risks borne by smaller firms, thereby sharpening rather than muting the contrast between small and large stocks \citep{leuz2003,bhattacharya2002,hail2006}. In that sense, formal legal development can cut both ways. It may ease financing frictions that otherwise inflate small-firm required returns, but it may also formalize market participation, impose scale-sensitive compliance burdens, and make size-related risk premia easier to detect in reported estimates.

The implication is that rule of law should not be treated here as a simple ``good institutions'' variable with a mechanically signed effect on the reported size slope \citep{beck2005b,hail2006,iliev2010,gao2009}. This ambiguity matters for the design: the formal institutional environment must be analyzed on its own terms before generalized trust is brought back in. If rule of law already performs part of the work that informal trust performs in weaker institutional settings, then the relevant hypothesis is not purely additive.

\subsection{Institutional Substitution and Testable Predictions}

The preceding subsections imply that trust and rule of law are not two interchangeable measures of institutional quality. They support exchange through different margins. Generalized trust can reduce the perceived danger of dealing with unfamiliar counterparties when information is soft and contracts are incomplete, while rule of law supplies formal enforcement, investor protection, and predictable legal recourse. If these functions overlap, the relevant prediction is conditional rather than simply additive: trust should matter most where formal enforcement is weaker, because in those settings informal assurance can partly substitute for legal infrastructure that investors and creditors otherwise lack.

This substitution logic has direct support in the trust-and-finance literature. \citet{guiso2004} show that social capital is especially strongly associated with financial development where legal enforcement is weak, while \citet{ahlerup2009} and \citet{bjrnskov2022} find that trust contributes more to growth and productivity in weaker institutional environments. The same mechanism is consistent with \citet{aghion2010}, who link distrust and regulation in a self-reinforcing equilibrium: where people expect opportunistic behavior, societies demand more formal control, and formal control can in turn crowd out trust. Applied to the size premium, this literature suggests that generalized trust may be most relevant for small, opaque firms precisely in countries where courts, disclosure enforcement, and investor protection provide less reliable assurance.

The hypothesis is nevertheless not that informal and formal institutions are always substitutes. Models of trust, regulation, and investment allow both substitution and complementarity, depending on the level of social capital and the credibility of legal enforcement \citep{carlin2009}. Empirical work on private credit also finds cases in which trust works through stronger formal institutions rather than replacing them \citep{cruzgarcia2019}, and experimental evidence points to complementarity between trust and contract enforcement in some exchange environments \citep{bartling2025}. More generally, social norms and legal enforcement can reinforce one another when laws are legitimate and broadly followed \citep{acemoglu2017}. These competing views rule out a mechanical prediction that more trust and better law must always move reported size coefficients in the same direction. The empirical question is instead whether the reported size-premium literature displays the particular conditional pattern implied by institutional substitution.

Two testable propositions follow from the substitution argument and the sign convention used in this paper: where rule of law is weak, higher generalized trust should be associated with less negative reported size slopes, and thus a weaker premium (more negative estimates imply a stronger one); and this trust association should decline as rule of law improves. In the interaction specification below, that corresponds to a positive trust association at low rule of law and a negative trust-by-rule-of-law interaction. Rule of law remains part of the empirical specification because it is the formal-institution anchor of the argument, but the literature does not imply a single directional prediction for its direct association with reported size slopes: stronger formal institutions can ease small-firm financing frictions, but they can also make small-firm risk, disclosure obligations, and fixed compliance costs more visible in market prices. We treat the substitution pattern as a sign-consistent prediction to be tested cautiously. The direct rule-of-law association is theoretically ambiguous: the friction-easing channel and the pricing-and-compliance channel imply opposite signs for the reported slope, so the estimated sign carries information about which channel dominates in the reported estimates.

\section{Data and Variables}\label{sec:data}

The estimated size premium varies widely across studies, countries, sample periods, and specifications. Using an estimate-level dataset that extends \citet{astakhov2019} with institutional and macro-financial measures, we ask which country-, study-, and model-level characteristics explain that variation, with particular attention to trust and rule of law. Table \ref{tab:data_sources} summarizes the data sources and the main variables.

\subsection{Size-Premium Estimates and Primary-Study Specifications}

The main dependent variable comes from primary asset-pricing studies that ask how stock returns vary with firm size. Each reported size coefficient becomes one observation in the meta-analysis. \citet{astakhov2019} report 1,746 such estimates from 102 published studies, plus a working-paper extension of 10 studies and 167 estimates, together with information on the study, sample period, country, empirical specification, estimation method, and reported precision. We start from that inventory, including the working-paper extension, use the reported coefficients as the outcome to be explained, and augment the estimate-level data with institutional and macro-financial variables.

In the underlying literature, firm size is usually measured as the market value of equity, commonly transformed by the natural logarithm because the cross-sectional distribution of firm values is highly skewed \citep{astakhov2019}. Because larger firms sit higher on the size scale, a more negative reported slope means that small firms earn relatively more than large firms. Less negative slopes therefore mean a weaker premium.

Primary studies estimate the size premium in two main ways. The first approach estimates a slope on firm size in a return regression, often in the Fama--MacBeth tradition and alongside market beta or other firm characteristics \citep{blume1970,fama1973}:
\begin{equation}
R_{kt} = \gamma_{0t} + \gamma_{1t}\widehat{\beta}_k + \gamma_{2t}\log(ME_{kt}) + Z_{kt}'\delta_t + u_{kt}.
\label{eq:primary_size_slope}
\end{equation}
Here \(k\) indexes firms and \(t\) indexes periods, \(ME_{kt}\) denotes market equity, and \(Z_{kt}\) collects any additional firm characteristics included by the primary study. The size coefficient, \(\gamma_{2t}\), or its time-series average across cross-sections is the object that enters the meta-analysis when it is reported with sufficient precision information.

The second approach estimates the size premium through sorted portfolios or a small-minus-big factor return \citep{fama1993}. We use reported regression slopes rather than raw small-minus-big premia (as in \citealp{astakhov2019}), because they retain the estimate-level standard errors needed for meta-regression, stay close to the marginal relation between returns and size, and depend less on portfolio-construction choices than raw factor or portfolio premia, which can also compress the within-sample variation that regression slopes retain \citep{lo1990,berk2000,fama2008}. The dependent variable is therefore the reported slope on firm size, not a small-firm-minus-large-firm return, and the sign is reversed relative to conventional premium language: stronger conventional size premia appear as more negative size slopes.

Following the source inventory, we use the reported coefficients in the units used by the primary studies rather than converting them into a common annualized effect size. Return definitions and horizons, and in some cases the construction of size, therefore vary across observations. Our dependent variable should accordingly be read as a reported-slope outcome, not as repeated measurements of one cardinal premium. The meta-regression conditions on the coded study-design and specification differences available in the inventory, and the illustrative magnitude in Section 7 is likewise expressed only in reported-slope space.

Keeping the evidence at the estimate level retains the heterogeneity of the literature, since studies differ in markets, periods, specifications, reporting choices, and estimation methods. It also creates a clear limit. The 1,613 rows are not independent observations; many come from the same study or the same country. For that reason, we report inference clustered separately by study and by country. The robustness section returns to the consequences of the highly unbalanced sample.

\subsection{Trust Measures}

To measure the trust environment around each estimate, we draw on the Integrated Values Surveys assembled from the European Values Study (EVS) Trend File 1981--2017 and the World Values Survey (WVS) Trend File 1981--2022 \citep{evs2022,haerpfer2022}. These dates give the outer span of the underlying surveys; the data do not form a balanced annual panel. EVS ends earlier than WVS, countries enter in different waves, fieldwork years are uneven, and not every trust question appears in every country-year survey. We therefore create a trust index for each observed country-year by averaging multiple trust questions, each taken from its nearest available survey year.

The baseline trust index is meant to capture measured generalized interpersonal trust rather than one exact survey wording. To reduce researcher discretion in index construction, it uses all available EVS/WVS questions that directly ask about trust in people or social groups, except for trust in family, which is kept for the placebo test below. It does not use trust or confidence in specific formal institutions, such as the army, police, or courts, because those items are closer to institutional confidence than to generalized interpersonal trust. The index should still be read as a cross-country proxy, since survey trust measures can raise comparability concerns across cultural settings \citep{reeskens2008}.

The resulting index has seven components. The broadest item is the standard generalized-trust question: ``Generally speaking, would you say that most people can be trusted or that you can't be too careful in dealing with people?'' The country-year measure is the share answering ``most people can be trusted.'' The other six components are country-year means of questions about trust in named groups. Five come from the more recent survey waves, which use the four-point prompt asking whether respondents trust the group ``completely, somewhat, not very much or not at all''; the groups are people in the respondent's neighborhood, people known personally, people met for the first time, people of another religion, and people of another nationality. The remaining item comes from earlier survey waves and asks whether respondents trust other people in their country on a five-point scale ranging from ``trust completely'' to ``not trust at all.''

The index construction follows the same logic for each country-year component. Non-substantive responses such as no answer, don't know, not applicable, and not asked are treated as missing. For the multi-category trust items, lower raw response values indicate greater trust, so the country-year means are reversed before standardization. The share answering that most people can be trusted already increases with trust and is not reversed. Each component is standardized as a z-score over the observed country-year survey cells: we subtract that component's mean and divide by its standard deviation, omitting missing cells from both calculations. The generalized-trust index is then the row mean of the available standardized components. Component coverage varies across country-year cells, and the components of a given cell can come from different survey years, so cells with fewer observed components measure generalized trust more noisily. A country-year receives a missing index only when none of the seven components is observed. Higher values therefore indicate more measured generalized trust.

Holding family trust out of the main index also gives a direct check on our interpretation. Trust inside the family is conceptually different from the broader nonfamily trust captured by the generalized-trust index, which spans acquaintances, neighbors, and strangers. The literature distinguishes generalized trust, which supports impersonal exchange, from particularized trust, which remains confined to family and close networks \citep{banfield1958,uslaner2002,delhey2011,alesina2010}. We therefore use family trust as a placebo: if the main pattern simply reflected the inclusion of any survey-based trust measure, family trust should produce a similar result. The placebo index uses two EVS/WVS items. The item from earlier survey waves asks how much respondents trust their family, with answers from ``trust them completely'' to ``do not trust them at all.'' The item from more recent survey waves uses the same named-group prompt as above for ``your family,'' with answers from ``trust completely'' to ``do not trust at all.'' These two country-year means are reversed where necessary, standardized, and averaged in the same way as the generalized-trust components. The robustness checks below show that the marginal effects of family trust do not reproduce the conditional substitution pattern found for generalized trust.

\subsection{Rule of Law and Other Institutional Measures}

Formal institutions are measured with the Worldwide Governance Indicators (WGI). Their rule-of-law series summarizes perceptions of contract enforcement, property rights, the courts, and the likelihood of crime and violence \citep{worldbank2025wgi,worldbank2025wgi_methodology,kaufmann2024wgi}. It is the natural counterpart to generalized trust in this paper because it maps most directly onto the formal enforcement mechanisms emphasized by law-and-finance and institutional-economics research \citep{north1990,laporta1998,laporta2006,hail2006}.

Each country-year observation is matched to the nearest available WGI year. The analysis also includes government effectiveness, regulatory quality, and control of corruption from the same source. Those measures return later as alternative formal-institution checks. Rule of law remains the baseline because it is closest to the legal enforcement and investor-protection channel in the hypothesis. For samples predating survey or WGI coverage, the nearest available value is necessarily measured later.

\subsection{Meta-Regression Moderators and Sample Construction}

Reported size slopes vary for reasons that are broader than the institutional mechanism studied in this paper. Primary studies use different return definitions, sample windows, trimming rules, estimation methods, portfolio constructions, control variables, publication outlets, and market environments. The meta-regression therefore treats the institutional variables as one candidate explanation among several. The moderators below follow the logic of the primary size-premium literature and of meta-analytic work on asset-pricing estimates: they describe how the estimate was produced, which return predictors the primary study allowed for, how precise the estimate is, how prominent the study is in the literature, and what market environment the estimate represents \citep{fama1992,fama1993,vandijk2011,astakhov2019,bajzik2020,irsova2023}.

One moderator family captures study design and data construction. Size estimates can differ in how the size variable and the sample are set up: studies define size relative to a sample average, cover different calendar periods, use longer or shorter samples, and trim extreme observations. They can also differ in how returns and the regression are specified, using excess rather than raw returns, individual stocks rather than portfolios, January-only returns, or a demeaned independent variable. These distinctions matter because the size premium has been linked to seasonality, microcap concentration, time variation, and the way noisy firm-level returns are aggregated \citep{banz1981,reinganum1981,keim1983,knez1997,vandijk2011,hou2019}. We code these features with indicator variables for the relevant design choices and continuous variables for the midpoint and length of the sample period.

A second family captures the primary-study estimation method. Fama--MacBeth regressions remain a standard tool for estimating average cross-sectional return relations because they run repeated cross-sections and then summarize the coefficient series over time \citep{fama1973,fama1992}. Other studies report pooled or ordinary least squares specifications. Methodological differences can affect both the level of the reported slope and its precision, especially when return shocks are correlated across firms or when standard errors are adjusted differently across studies \citep{astakhov2019}. The meta-regression therefore includes indicators for Fama--MacBeth and ordinary least squares estimation.

A third family records the asset-pricing controls included in the primary specification. This block is central because the size premium is partly entangled with other firm characteristics. Book-to-market, market beta, momentum, profitability, investment-related characteristics, leverage, and liquidity can each absorb part of what would otherwise appear as a size slope. So can volatility, price levels, information variables, ownership, growth, distress, and additional size controls \citep{basu1983,bhandari1988,fama1992,fama1993,jegadeesh1993,carhart1997,amihud2002,dichev1998,fama2015,asness2018,hou2019}. The analysis codes these controls as broad indicator groups, so the main text refers to the economic family rather than to every underlying variable name. The detailed grouping is reported in Appendix Table \ref{tab:control_dictionary}.

A fourth family covers publication and precision context: \citet{astakhov2019} identify publication bias in the size-premium literature, and publication bias in meta-analysis is commonly diagnosed through a systematic relation between a reported estimate and its standard error \citep{stanley2005,stanley2008}. In the absence of such selection, sampling imprecision should not by itself predict the reported coefficient once the underlying effect and study design are accounted for. The meta-regression therefore includes the reported standard error as a precision moderator. Publication outlet and citation counts are included for a different reason: highly visible studies and more cited estimates may use more standardized designs, may be more likely to report statistically significant effects, or may come from literatures in which certain specifications became canonical. We therefore control for reported precision, journal impact, and citations without interpreting them as causal determinants of the true size premium.

Finally, size effects may differ across macro-financial environments, including interest-rate conditions, aggregate market volatility, financial development, income levels, and growth conditions. These variables are also relevant for the institutional interpretation because financial development and macroeconomic conditions affect the financing constraints and risk exposures of small firms \citep{beck2005b,levine2005,hail2006,beck2008}. The baseline specification therefore includes bond yields, market-return volatility, private credit to GDP, GDP per capita, and GDP growth as country-period controls.

Some continuous measures enter the regression after simple transformations that make their scales easier to compare. The midpoint of each study's sample period is expressed relative to the sample average. Sample length and citation counts are logged because they are highly skewed. GDP per capita is also logged, private credit is expressed as a ratio rather than a percentage, and market volatility and GDP growth are multiplied by 100 so that they enter in percentage-point units. Variable and control-group definitions are reported in Appendix Tables \ref{tab:variable_dictionary} and \ref{tab:control_dictionary}; the main text groups them by the economic reason for including them.

The final sample is the part of the collected literature that can be linked cleanly to the institutional and control information needed for the baseline specification. Our starting pool includes \citet{astakhov2019}'s published-study inventory and its working-paper extension (the count reconciliation is in Appendix A). Because trust and rule of law are measured at the country level, we drop estimates that pool several countries together. We also exclude estimates with sample midpoints before 1960, observations with missing institutional or control information, and observations that become non-finite after the log transformations. The complete-case analysis sample contains 1,613 estimates from 105 studies and 31 countries. It is highly unbalanced: the United States accounts for 1,115 observations, or 69.1\% of the sample. The design remains useful for studying reported estimates, although this imbalance precludes a clean cross-country causal interpretation. Appendix Table \ref{tab:final_study_counts} reports the source-study counts in the final analysis sample.

Reported size slopes and standard errors have long tails, typical of primary-study estimates pooled across papers. The baseline specifications therefore apply mild 1/99 winsorization to these two variables after the complete-case sample is formed. Table \ref{tab:summary_stats} reports the unwinsorized distribution of the analysis sample. The empirical strategy and robustness sections describe the tail treatment and its checks.

\begin{table}[t]
\centering
\caption{Data Sources and Main Variables}
\label{tab:data_sources}
\begin{threeparttable}
\small
\begin{tabular}{p{0.22\textwidth}p{0.24\textwidth}p{0.26\textwidth}r}
\toprule
Object & Source & Construction and role & Coverage \\
\midrule
Reported size coefficients & \citet{astakhov2019} & Reported slopes from regressions of returns on firm size; dependent variable & 1,613 \\
Generalized trust index & EVS/WVS & Row mean of seven standardized generalized-trust components; informal-institution moderator & 1,613 \\
Rule of law & WGI & Nearest-year country match to WGI rule-of-law estimate; formal-institution moderator & 1,613 \\
Study and estimate controls & \citet{astakhov2019} & Design, specification, precision, and publication-context controls & 1,613 \\
Macro-financial controls & Country-year controls linked to the \citet{astakhov2019} estimates & Market and macroeconomic conditions, transformed as in the preferred specification & 1,613 \\
Family-trust placebo & EVS/WVS & Row mean of standardized family-trust items; placebo institutional index & 1,613 \\
\bottomrule
\end{tabular}
\begin{tablenotes}
\footnotesize
\item Notes: Coverage is measured in the complete-case analysis sample used in the baseline specification. EVS/WVS denotes European Values Study/World Values Survey; WGI denotes Worldwide Governance Indicators.
\end{tablenotes}
\end{threeparttable}
\end{table}

\begin{table}[t]
\centering
\caption{Summary Statistics in the Analysis Sample}
\label{tab:summary_stats}
\begin{threeparttable}
\begin{tabular}{lrrrrrrr}
\toprule
Variable & Mean & SD & Min. & P25 & Median & P75 & Max. \\
\midrule
Size slope & -0.091 & 0.472 & -3.600 & -0.101 & -0.005 & -0.000 & 4.447 \\
Standard error & 0.089 & 0.278 & 0.000 & 0.001 & 0.023 & 0.054 & 3.789 \\
Generalized trust index & 0.762 & 0.504 & -1.001 & 0.900 & 0.900 & 0.993 & 1.886 \\
Rule of law & 1.407 & 0.413 & -1.290 & 1.500 & 1.500 & 1.500 & 1.951 \\
Family-trust placebo index & 0.045 & 0.269 & -1.152 & -0.039 & -0.039 & -0.039 & 1.208 \\
\bottomrule
\end{tabular}
\begin{tablenotes}
\footnotesize
\item Notes: The table reports the unwinsorized distribution of the 1,613-observation complete-case analysis sample. The sample contains 105 studies and 31 countries; 1,115 observations, or 69.1\%, are from the United States. Because of this weight, the estimate-level quantiles of the institutional variables are pinned at the U.S. values; the country-level quantiles used for interpretation are reported in Appendix Table \ref{tab:institutional_quantiles_appendix}. The statistics use the same transformations as the baseline interaction specification. Values are shown to three decimals; the 75th percentile of the size slope is negative but smaller than 0.0005 in magnitude.
\end{tablenotes}
\end{threeparttable}
\end{table}

\section{Empirical Strategy}
\label{sec:empirical-strategy}

\subsection{Baseline Meta-Regression}

We ask whether trust and rule of law retain a systematic association with reported size slopes once the other sources of meta-analytic heterogeneity are held constant.

The specification follows directly from the conditional-substitution argument developed above. If generalized trust partly substitutes for weak formal enforcement, its association with the reported size slope should depend on rule of law. A purely additive specification would miss that possibility, so the baseline includes trust, rule of law, and their interaction. The institutional coefficients are estimated from cross-country and over-time variation in the matched institutional measures, with study and specification differences entering through the controls; the coefficients describe conditional associations in the literature.

The baseline estimating equation is:

\begin{equation}
\sizeslope_{ijc}
= \alpha
+ \beta_1 \trust_{ijc}
+ \beta_2 \rol_{ijc}
+ \beta_3 \left(\trust_{ijc} \times \rol_{ijc}\right)
+ X_{ijc}'\gamma
+ \varepsilon_{ijc}.
\label{eq:baseline}
\end{equation}

In Equation \eqref{eq:baseline}, \(i\) indexes an estimate, \(j\) indexes the study from which the estimate is drawn, and \(c\) indexes the country to which the estimate refers. The dependent variable, \(\sizeslope_{ijc}\), is the reported size coefficient collected by \citet{astakhov2019} (the \(\gamma_{2t}\) of Equation \eqref{eq:primary_size_slope} or its time-series average for Fama--MacBeth estimates, and the analogous size slope for pooled and least-squares specifications). Because the underlying coefficient is a slope on firm size, more negative values imply a stronger conventional small-minus-big premium.

The variables \(\trust_{ijc}\) and \(\rol_{ijc}\) denote the generalized-trust index and the rule-of-law measure assigned to the estimate's country and sample timing. They are left uncentered in the preferred specification. Here \(\beta_1\) is the trust association when rule of law is zero, and \(\beta_2\) the rule-of-law association when trust is zero. These are useful anchors, but the economically meaningful quantities are the marginal effects over the observed support.

The vector \(X_{ijc}\) collects the remaining sources of meta-analytic heterogeneity discussed in the data section. These include study-design indicators, asset-pricing specification choices, estimate precision, journal impact and citation measures, market characteristics, and macro-financial controls. We include the reported standard error as the precision moderator used in funnel-asymmetry diagnostics \citep{stanley2005,stanley2008}, together with journal impact and citations. We do not fit a full selection model, because the target is the institutional moderators of reported slopes, not the bias-corrected mean size premium. The baseline estimator is unweighted ordinary least squares.

Before estimation, the reported size slope and its standard error are winsorized at the 1st and 99th percentiles within the final complete-case model frame.\footnote{In the 1,613-observation model frame, 1/99 winsorization clips 17 observations in the lower tail and 16 observations in the upper tail of each variable. For the reported size slope, the support changes from \([-3.600,4.447]\) to \([-1.735,1.330]\), while the mean and standard deviation change from \(-0.091\) and \(0.472\) to \(-0.087\) and \(0.344\). For the standard error, the support changes from \([3.05\times 10^{-7},3.789]\) to \([0.000123,1.614]\), while the mean and standard deviation change from \(0.089\) and \(0.278\) to \(0.084\) and \(0.226\).} This mild tail treatment limits the leverage of extreme reported coefficients and imprecise estimates while leaving the analysis sample and the institutional variables unchanged. Since many estimates come from the same studies and countries, the results section reports the same point estimates with two cluster-robust covariance matrices: one clustered by study and one clustered by country. This presentation keeps the identifying variation transparent and avoids treating the 1,613 estimate-level observations as independent country-level evidence.

\subsection{Marginal-Effect Definitions}

Because the baseline model contains an interaction, the raw main-effect coefficients do not by themselves answer the substantive question. The coefficient on trust is evaluated at a zero value of rule of law, and the coefficient on rule of law is evaluated at a zero value of the trust index. Those values are interpretable reference points for an uncentered specification, but the economic question is how the trust association evolves as formal legal quality changes over the observed support of the data.

For this reason, the empirical interpretation is based on marginal effects. The first marginal effect asks how the reported size slope changes with trust at a given level of rule of law. The second asks how it changes with rule of law at a given level of trust.

\begin{align}
\frac{\partial \sizeslope}{\partial \trust}
&= \beta_1 + \beta_3 \rol, \label{eq:me_trust}\\
\frac{\partial \sizeslope}{\partial \rol}
&= \beta_2 + \beta_3 \trust. \label{eq:me_rol}
\end{align}

For each marginal effect, uncertainty is computed from the covariance matrix of the relevant linear combination of coefficients. The plotted confidence bands use the same study-clustered and country-clustered covariance matrices as the OLS tables. The figures also show the observed support of the conditioning variable, so the reader can distinguish parts of the interaction surface that are well represented in the data from extrapolated regions.

\subsection{Model Uncertainty}

Meta-regressions face substantial model uncertainty because the literature supplies many plausible controls and only limited theory about which study-design variables must appear in every specification. A single full-control OLS regression is transparent, but it can give a misleading sense of precision if the institutional pattern depends on a particular control set. To address this concern, the analysis also uses Bayesian model averaging (BMA), a standard tool in economic meta-analysis \citep{raftery1997,eicher2011,zeugner2015}.

The BMA exercise keeps the dependent variable, institutional variables, transformations, and tail treatment aligned with the baseline OLS specification. It samples models over the control variables in \(X_{ijc}\), while forcing generalized trust, rule of law, and their interaction to appear in every sampled model. We fix the institutional block deliberately. The design does not ask whether trust and rule of law survive selection against a menu of unrelated controls; it asks whether their sign pattern is sensitive to uncertainty about the other moderators. Consequently, posterior inclusion probabilities (PIP) of one for the institutional terms are mechanical and should not be read as evidence that BMA selected those terms from the data.

The baseline BMA uses the BRIC \(g\)-prior, a random model-size prior, and birth-death Markov chain Monte Carlo. The reported BMA uses 100,000 burn-in draws, 200,000 retained iterations, and is configured to store up to 5,000 best models. Coefficients are summarized as posterior means and posterior intervals after averaging across the sampled model space. For interaction interpretation, marginal-effect intervals are computed from posterior covariance terms so that the uncertainty in linear combinations such as \(\beta_1 + \beta_3\rol\) is treated consistently.\footnote{The standard output of the \texttt{BMS} package reports posterior means and marginal posterior standard deviations, but it does not report the cross-coefficient posterior covariances needed for interaction marginal effects. The BMA calculation therefore retains posterior cross-moments for the institutional coefficient block during the Markov chain. The resulting summaries were checked against the package's native output before using the retained covariance terms to form the marginal-effect intervals.}

The BMA results are used as model-uncertainty evidence, not as a substitute for clustered OLS inference. Their uncertainty bands are Bayesian posterior intervals under the specified priors and model space; they are not study-clustered or country-clustered standard errors. They also do not resolve the observational nature of the design, the limited institutional variation, or the sample-composition issues examined later. Their role is narrower. They show whether the preferred pattern is qualitatively stable when the large set of study-design and macro controls is not treated as fixed with certainty.

\section{Results}

\subsection{Descriptive Heterogeneity}

\begin{figure}[!t]
\centering
\begin{minipage}{0.86\textwidth}
\centering
\IfFileExists{Figures/size_slope_country_boxplot.pdf}{%
  \includegraphics[width=\textwidth,height=0.48\textheight,keepaspectratio]{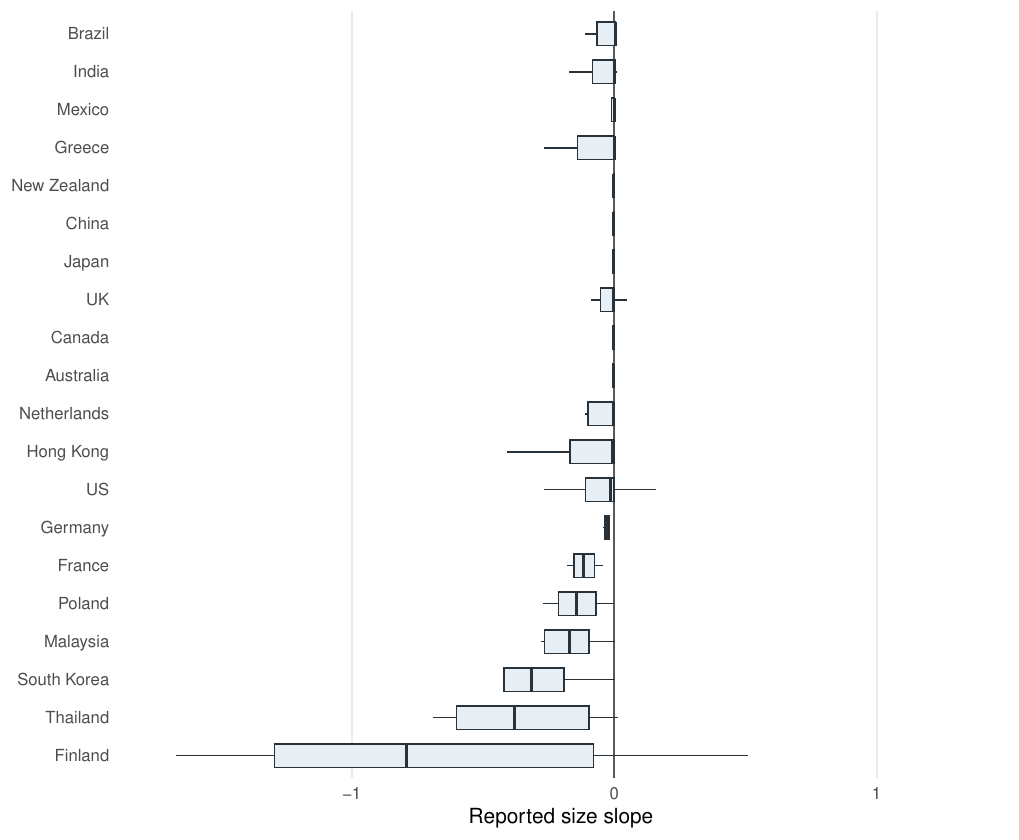}%
}{%
  \fbox{\parbox[c][1.2in][c]{\textwidth}{\centering Placeholder for country-level box plot of reported size slopes.}}%
}
\caption{Reported Size Slopes Vary Widely across Countries}
\label{fig:size_country_boxplot}
\par\medskip
\noindent\parbox[t]{\textwidth}{\footnotesize Notes: The figure shows the distribution of 1/99 winsorized reported size slopes in the preferred complete-case sample for countries with at least five estimates. Countries are ordered by their median winsorized size slope. Outlier points are not drawn, but all country observations enter the box summaries. More negative slopes imply larger conventional size premia.}
\end{minipage}
\end{figure}

Reported size slopes differ sharply across countries (Figure \ref{fig:size_country_boxplot}), which is why a heterogeneity analysis is warranted. The analysis sample contains 1,613 reported size slopes from 105 studies and 31 countries. In the unwinsorized sample, the mean reported slope is \(-0.091\), with a standard deviation of 0.472. The 1/99 winsorized outcome used in the baseline regression ranges from about \(-1.735\) to 1.330. These numbers are large relative to the average estimate and indicate that the literature contains much more than small sampling fluctuations around a single common size effect.

The dispersion is economically meaningful: a more negative reported coefficient means a stronger size premium. The cross-country box plot should nevertheless be read descriptively. The sample is highly unbalanced: the United States contributes 1,115 observations, or 69.1\% of the analysis sample. The figure therefore motivates the meta-regression but does not by itself establish a country-level institutional relationship.

\subsection{Baseline Institutional Meta-Regression}

\begin{table}[t]
\centering
\caption{Baseline Institutional Meta-Regression: Rule of Law as the More Stable Association}
\label{tab:baseline}
\begin{threeparttable}
\small
\begin{tabular}{lrrrrrr}
\toprule
 & \multicolumn{3}{c}{Study-clustered} & \multicolumn{3}{c}{Country-clustered} \\
\cmidrule(lr){2-4}\cmidrule(lr){5-7}
 & Estimate & SE & \(p\)-value & Estimate & SE & \(p\)-value \\
\midrule
Generalized trust & 0.146 & 0.086 & 0.091 & 0.146 & 0.112 & 0.192 \\
Rule of law & -0.168 & 0.077 & 0.028 & -0.168 & 0.070 & 0.016 \\
Generalized trust $\times$ rule of law & -0.108 & 0.073 & 0.138 & -0.108 & 0.073 & 0.139 \\
\midrule
Controls & \multicolumn{3}{c}{Yes} & \multicolumn{3}{c}{Yes} \\
Observations & \multicolumn{3}{c}{1,613} & \multicolumn{3}{c}{1,613} \\
Studies & \multicolumn{3}{c}{105} & \multicolumn{3}{c}{105} \\
Countries & \multicolumn{3}{c}{31} & \multicolumn{3}{c}{31} \\
\bottomrule
\end{tabular}
\begin{tablenotes}
\footnotesize
\item Notes: The dependent variable is the 1/99 winsorized reported size slope. Winsorization is applied after forming the complete-case model frame. The table reports the same OLS point estimates with alternative cluster-robust covariance matrices. Controls include the study-design, asset-pricing specification, precision, publication-context, market, and macro-financial moderators described above. See Appendix Table \ref{tab:full_baseline_ols} for the full set of results. More negative reported slopes imply larger conventional size premia.
\end{tablenotes}
\end{threeparttable}
\end{table}

The clearest pattern in the data is that stronger rule of law goes with a larger conventional size premium across much of the relevant trust support (Table \ref{tab:baseline}): the rule-of-law coefficient is negative and conventionally informative under both study- and country-clustered inference. This is consistent with formalization, disclosure, and risk-pricing channels, and with the possibility that better legal systems make small-firm risks more visible.

The trust interaction corresponds to the substitution hypothesis. Its negative sign means that the trust association declines as rule of law rises. The marginal effect of trust combines the trust coefficient with the interaction at each level of rule of law, so Figure \ref{fig:marginal_effects} provides the substantive interpretation. The interaction coefficient itself is imprecise: the study-clustered \(p\)-value is 0.138, and the country-clustered \(p\)-value is 0.139.

We treat the institutional variables as a block, not as separable causes: trust, rule of law, and broad development move together empirically, so the baseline interaction should not be read as a clean decomposition of independent institutional causes.

\subsection{Marginal Effects and Interpretation}

\begin{figure}[t]
\centering
\begin{subfigure}{0.48\textwidth}
\centering
\IfFileExists{Figures/trust_marginal_effect_plot.pdf}{%
  \includegraphics[width=\textwidth]{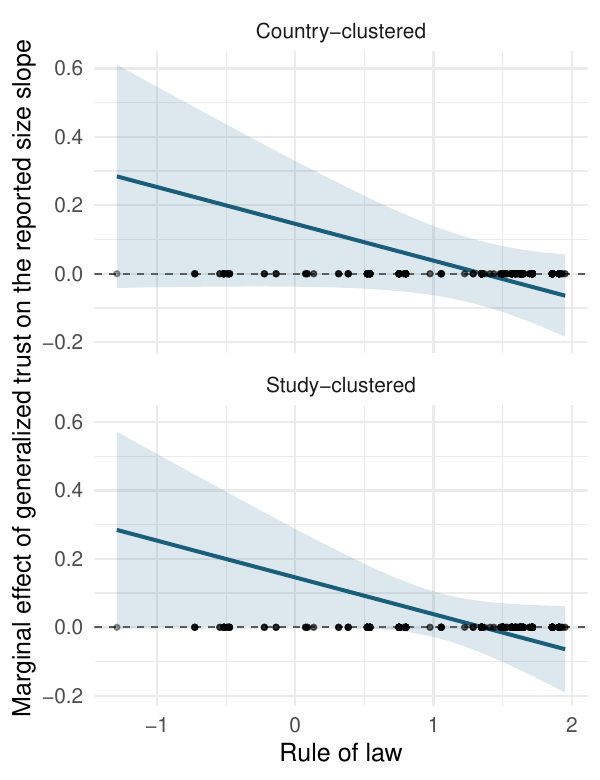}%
}{%
  \fbox{\parbox[c][1.2in][c]{\textwidth}{\centering Placeholder for marginal effect of trust over rule of law.}}%
}
\caption{Marginal effect of generalized trust over rule of law}
\end{subfigure}
\hfill
\begin{subfigure}{0.48\textwidth}
\centering
\IfFileExists{Figures/rol_marginal_effect_plot.pdf}{%
  \includegraphics[width=\textwidth]{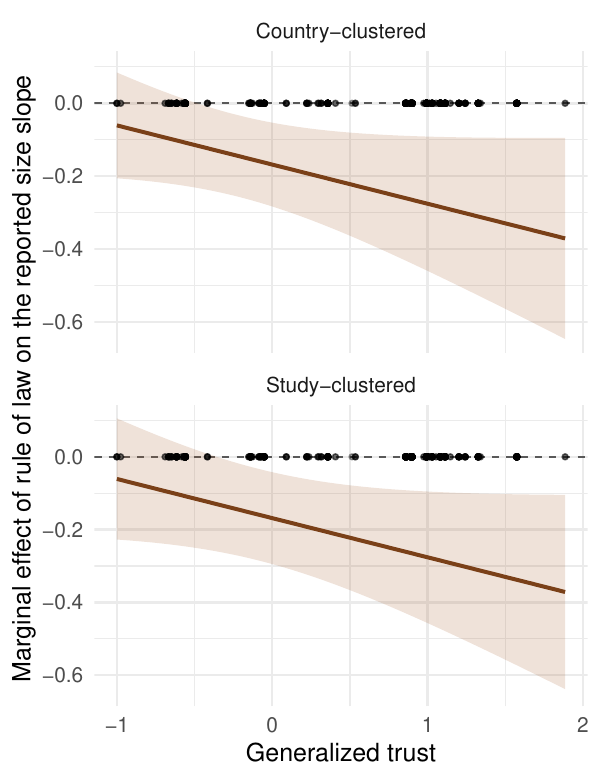}%
}{%
  \fbox{\parbox[c][1.2in][c]{\textwidth}{\centering Placeholder for marginal effect of rule of law over trust.}}%
}
\caption{Marginal effect of rule of law over generalized trust}
\end{subfigure}
\caption{Marginal Effects, Baseline OLS: the Estimated Trust Association Declines with Rule of Law}
\label{fig:marginal_effects}
\par\medskip
\noindent\parbox[t]{\textwidth}{\footnotesize Notes: The figures report marginal effects implied by the baseline OLS interaction model. Bands are 90\% confidence intervals computed from the relevant clustered covariance matrix. Support points are plotted on the zero line. Positive effects increase the reported size slope and therefore imply weaker conventional size premia; negative effects imply stronger conventional size premia.}
\end{figure}

Figure \ref{fig:marginal_effects} translates the interaction into the two conditional associations that matter for the economic interpretation. Panel (a) shows that the marginal effect of generalized trust on the reported size slope is largest in weaker-rule-of-law settings and declines as rule of law improves. At average or median rule of law, the implied trust effect is close to zero and is not statistically distinguishable from zero. The positive trust association is therefore not an average-sample result. At the plotted 90\% confidence level, it is positive only over low rule-of-law values, approximately from \(-1.00\) to 0.42 under study-clustered inference, and no part of the country-clustered band lies entirely above zero. The figure is consistent with higher trust going with a smaller premium where formal legal institutions are weaker, but that part of the evidence is narrow and sensitive to the clustering dimension.

\begin{figure}[!t]
\centering
\IfFileExists{Figures/institutional_partial_effect_heatmap.pdf}{%
  \includegraphics[width=0.78\textwidth]{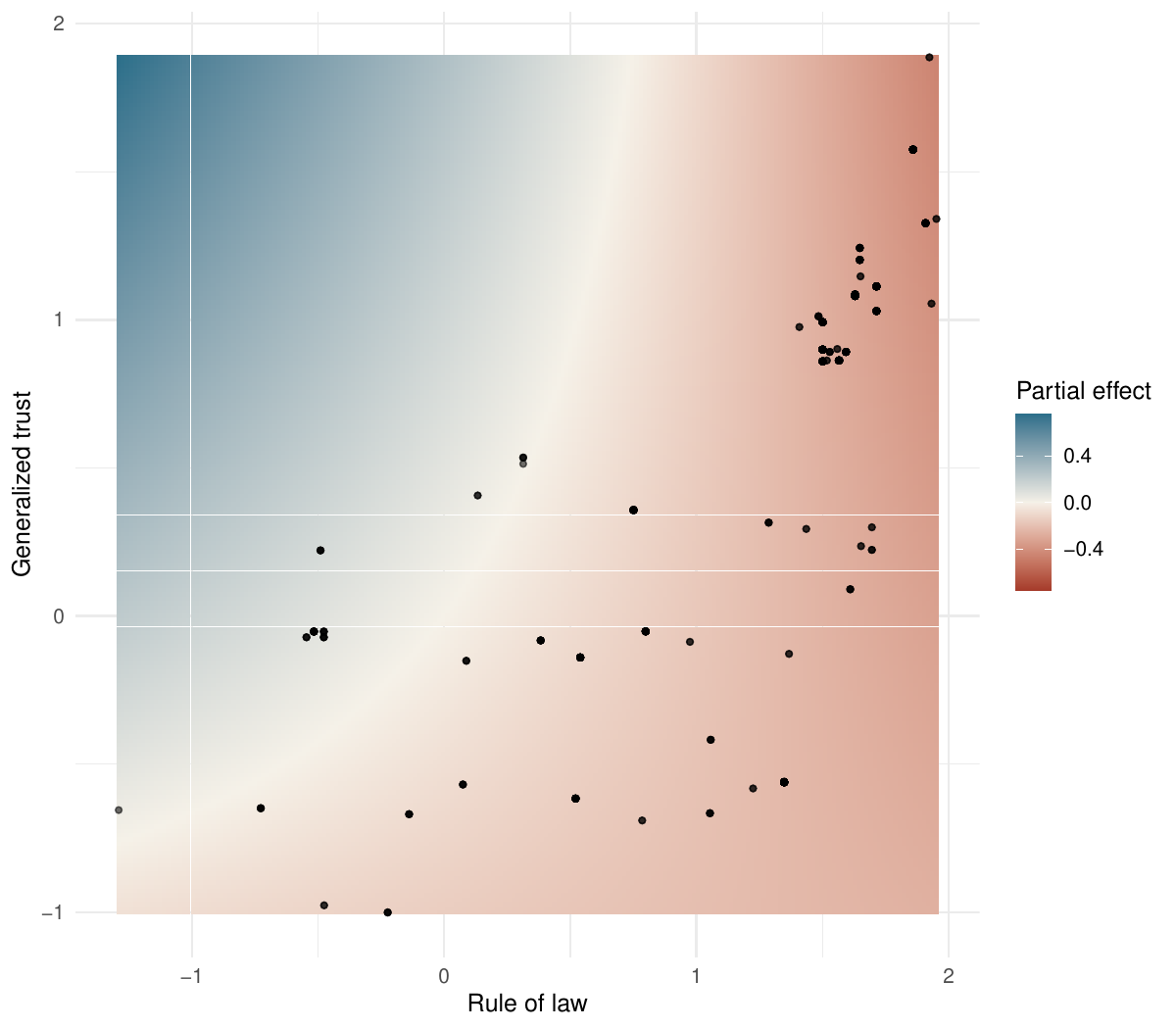}%
}{%
  \fbox{\parbox[c][1.2in][c]{0.78\textwidth}{\centering Placeholder for institutional partial-effect surface.}}%
}
\caption{The Fitted Institutional Contribution Is Most Negative Where Rule of Law Is High}
\label{fig:surface}
\par\medskip
\noindent\parbox[t]{\textwidth}{\footnotesize Notes: The figure isolates the fitted contribution of the generalized-trust, rule-of-law, and interaction terms from the baseline OLS regression. It is not the total fitted size slope. Points show observed combinations of trust and rule of law.}
\end{figure}

Panel (b) shows the complementary marginal effect of rule of law. This association is more consistently negative. The rule-of-law effect becomes more negative as generalized trust rises, and the 90\% confidence band excludes zero over much of the observed trust support: approximately from \(-0.35\) to 1.89 with study-clustered inference and from \(-0.48\) to 1.89 with country-clustered inference. Because negative effects make the reported size slope more negative, this pattern implies a stronger conventional size premium in higher-rule-of-law environments over most of the relevant trust range. This finding is not the simple claim that all better institutions reduce the size premium. Rather, the evidence suggests that generalized trust and formal legal quality enter the reported size-premium literature in a conditional way: the trust association is concentrated in weak-formal-institution settings, while the rule-of-law association is more visible across the main support of the trust index.

Figure \ref{fig:surface} summarizes the same pattern as a joint institutional surface, holding the other controls fixed. The lowest fitted institutional contributions appear where rule of law is high and trust is moderate or high, which corresponds to more negative reported size slopes and hence larger conventional size premia. The positive trust contribution is concentrated in weaker-rule-of-law settings and declines as the formal legal environment improves. The plot displays the interaction over the observed support; it does not map the total size premium in each country.
\subsection{Bayesian Model Averaging}

\begin{table}[!h]
\centering
\caption{Bayesian Model Averaging: Posterior Signs Match the Baseline OLS Pattern}
\label{tab:bma_institutional}
\begin{threeparttable}
\small
\begin{tabular}{lrrrr}
\toprule
Variable & Mean & SD & 90\% low & 90\% high \\
\midrule
Rule of law & -0.126 & 0.046 & -0.202 & -0.050 \\
Generalized trust & 0.135 & 0.066 & 0.026 & 0.244 \\
Generalized trust $\times$ rule of law & -0.092 & 0.043 & -0.162 & -0.021 \\
\bottomrule
\end{tabular}
\begin{tablenotes}
\footnotesize
\item Notes: The table reports primary MCMC-frequency BMA posterior summaries. The institutional variables are forced into every sampled model, so their posterior inclusion probabilities equal one by design and are not reported as selection evidence. Posterior intervals are Bayesian model-averaged intervals, not cluster-robust confidence intervals.
\end{tablenotes}
\end{threeparttable}
\end{table}

Table \ref{tab:bma_institutional} reports the Bayesian model-averaging counterpart to the baseline OLS specification. The exercise varies the study-design, specification, precision, publication-context, market, and macro-financial controls included in the meta-regression while keeping the dependent and institutional variables fixed. It assesses whether the institutional sign pattern persists under control-set uncertainty.

\begin{figure}[t]
\centering
\begin{subfigure}{0.48\textwidth}
\centering
\IfFileExists{Figures/bma_trust_marginal_effect_plot.pdf}{%
  \includegraphics[width=\textwidth]{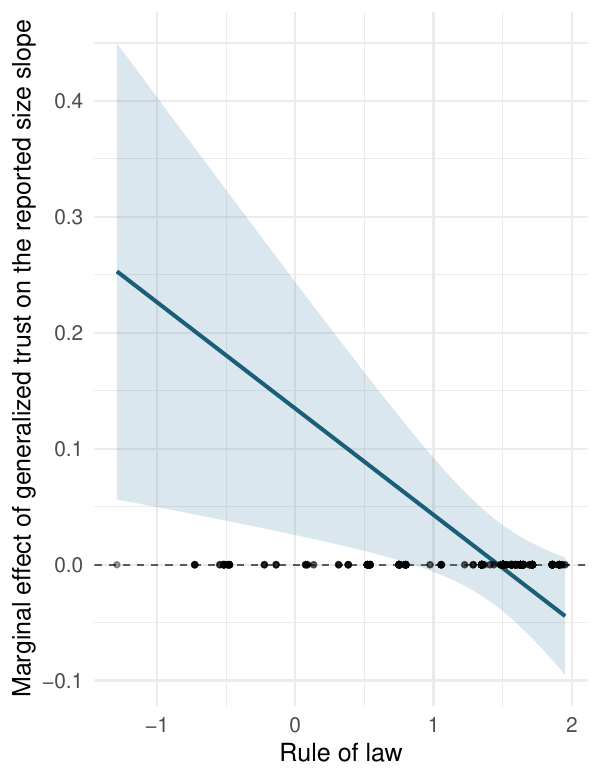}%
}{%
  \fbox{\parbox[c][1.2in][c]{\textwidth}{\centering Placeholder for BMA marginal effect of trust over rule of law.}}%
}
\caption{Marginal effect of generalized trust over rule of law}
\end{subfigure}
\hfill
\begin{subfigure}{0.48\textwidth}
\centering
\IfFileExists{Figures/bma_rol_marginal_effect_plot.pdf}{%
  \includegraphics[width=\textwidth]{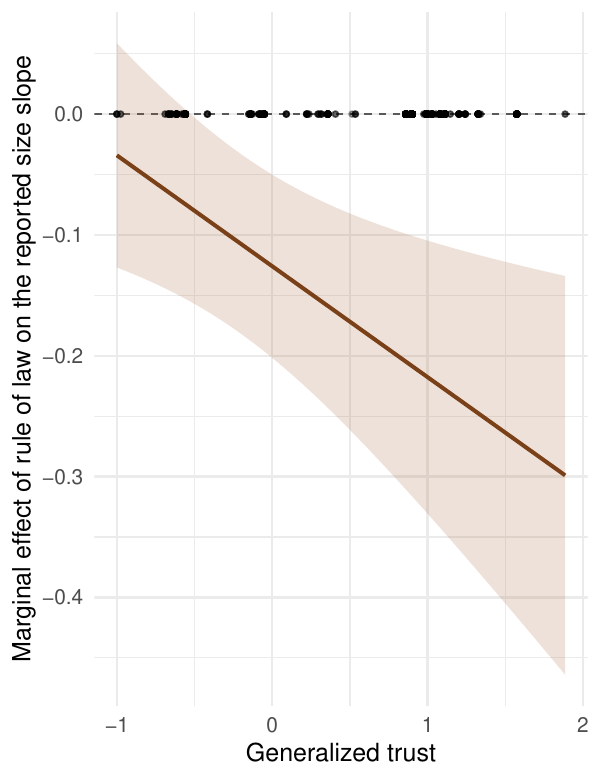}%
}{%
  \fbox{\parbox[c][1.2in][c]{\textwidth}{\centering Placeholder for BMA marginal effect of rule of law over trust.}}%
}
\caption{Marginal effect of rule of law over generalized trust}
\end{subfigure}
\caption{Marginal Effects, Bayesian Model Averaging: the Sign Pattern Persists under Control-Set Uncertainty}
\label{fig:bma_marginal_effects}
\par\medskip
\noindent\parbox[t]{\textwidth}{\footnotesize Notes: The figures report posterior marginal effects implied by the baseline Bayesian model-averaging interaction specification. Bands are 90\% posterior intervals computed from model-averaged posterior covariance terms. Support points are plotted on the zero line. Positive effects increase the reported size slope and therefore imply weaker conventional size premia; negative effects imply stronger conventional size premia.}
\end{figure}

The BMA posterior means carry the same signs as the OLS point estimates: positive for trust, negative for rule of law and for the interaction. The 90\% posterior intervals for all three institutional terms exclude zero under the reported BMA specification. Because these are model-averaged posterior intervals that treat the 1,613 estimates as independent observations, they leave the clustered-inference limitations unresolved. Figure \ref{fig:bma_marginal_effects} reports the corresponding posterior marginal effects. The posterior trust marginal effect is positive over lower rule-of-law values, approximately from \(-1.29\) to 0.84, and fades around the high-rule-of-law mass of the sample. The posterior rule-of-law marginal effect is negative over much of the trust support, approximately from \(-0.52\) to 1.89. Appendix Table \ref{tab:full_baseline_ols} reports the full OLS and BMA coefficient summaries, and Appendix Figure \ref{fig:bma_coefficients_appendix} shows the broader model-averaged coefficient pattern.

Stepping back from the BMA exercise, the remainder of this section summarizes the results as a whole. Reported size-premium estimates are not organized by a simple additive story in which all favorable institutions move the premium in the same direction. Rule of law is the more stable finding: its association with the reported size slope holds under both clusterings. The trust finding is less stable: trust is positively associated with the slope mainly where rule of law is low, which implies a weaker premium in weaker formal-institutional settings. The BMA exercise is useful because it preserves this sign pattern under control-set uncertainty; it does not remove the sample-dependence and clustering uncertainty that remain visible in OLS.

The full results in Appendix Table \ref{tab:full_baseline_ols} also show that the institutional terms are not the only systematic sources of heterogeneity. The most stable non-institutional patterns are concentrated in study design, sample construction, and primary-study specification choices. Estimates based on individual-stock samples and January-return specifications are associated with more negative reported size slopes, which corresponds to larger conventional size premia. Momentum controls are also associated with more negative reported size slopes. By contrast, trimming, value controls, price controls, and higher GDP per capita are associated with less negative reported slopes, implying smaller conventional size premia conditional on the other moderators. These variables are the clearest cases where OLS significance and high BMA posterior inclusion probabilities point in the same direction. Market-characteristic controls also have a strong positive OLS association, but their BMA evidence is less decisive once the rest of the control set is treated as uncertain.

\section{Robustness and Diagnostics}

The preceding section reports a rule-of-law association and a conditional trust pattern in reported size-premium estimates. This section asks how far that pattern changes when the analysis alters the measurement of trust and formal institutions, the control set, the tail treatment, the weighting and inference choices, and the composition of the estimate-level sample. Because the baseline specification applies mild 1/99 winsorization to the reported size slope and standard error, Table \ref{tab:robustness} includes both the no-winsorization benchmark and the more aggressive 5/95 stress test.

\begin{table}[!t]
\centering
\caption{Robustness and Diagnostics: Rule of Law Is the More Stable of the Two Institutional Terms}
\label{tab:robustness}
\begin{threeparttable}
\scriptsize
\setlength{\tabcolsep}{3pt}
\begin{tabular}{
  >{\raggedright\arraybackslash}p{0.21\textwidth}
  >{\raggedright\arraybackslash}p{0.1\textwidth}
  >{\raggedright\arraybackslash}p{0.1\textwidth}
  >{\raggedright\arraybackslash}p{0.1\textwidth}
  >{\raggedright\arraybackslash}p{0.31\textwidth}
}
\toprule
Check & Trust/focal \newline coef. & Formal-institution coef. & Interaction \newline coef. & Main reading \\
\midrule
Flat-data OLS & 0.173 \newline \((0.114)\) & -0.196 \newline \((0.018)\) & -0.121 \newline \((0.061)\) & Larger interaction when trust is treated as persistent \\
Residualized trust & 0.150 \newline \((0.133)\) & -0.031 \newline \((0.699)\) & -0.153 \newline \((0.051)\) & Larger interaction after removing linear overlap \\
Reduced-control OLS & 0.058 \newline \((0.306)\) & -0.110 \newline \((0.043)\) & -0.043 \newline \((0.284)\) & Same signs using high-PIP controls only \\
No winsorization OLS & 0.161 \newline \((0.147)\) & -0.183 \newline \((0.011)\) & -0.104 \newline \((0.150)\) & Pattern is not created by tail treatment \\
5/95 winsorized OLS & 0.058 \newline \((0.149)\) & -0.056 \newline \((0.018)\) & -0.036 \newline \((0.158)\) & Aggressive compression attenuates the result \\
Precision-weighted limiting diagnostic & -0.001 \newline \((0.516)\) & 0.000 \newline \((0.906)\) & 0.001 \newline \((0.250)\) & Coefficients collapse under concentrated weights \\
Family-trust placebo & 0.041 \newline \((0.462)\) & -0.134 \newline \((0.021)\) & -0.059 \newline \((0.101)\) & Marginal effects do not reproduce the conditional substitution pattern \\
Regulatory-quality substitute & -0.053 \newline \((0.660)\) & 0.078 \newline \((0.115)\) & -0.030 \newline \((0.700)\) & No comparable association \\
Government-effectiveness substitute & -0.024 \newline \((0.821)\) & 0.020 \newline \((0.738)\) & -0.031 \newline \((0.673)\) & No comparable association \\
Control-of-corruption substitute & -0.054 \newline \((0.645)\) & 0.145 \newline \((0.307)\) & -0.051 \newline \((0.378)\) & No comparable association \\
Low-rule-of-law median split, 1/99 winsorized & -0.089 \newline \((0.338)\) & 0.087 \newline \((0.189)\) & 0.078 \newline \((0.479)\) & Weak under baseline tail treatment \\
Low-rule-of-law median split, 5/95 winsorized & -0.064 \newline \((0.036)\) & 0.007 \newline \((0.726)\) & 0.075 \newline \((0.034)\) & Stronger only under aggressive tail treatment \\
Non-U.S. OLS & 0.036 \newline \((0.448)\) & -0.058 \newline \((0.039)\) & -0.011 \newline \((0.798)\) & Continuous interaction collapses \\
Collapsed country and study-country evidence & -- & -- & -0.027 \newline \((0.637)\); \newline -0.091 \newline \((0.183)\) & No strong collapsed cross-country relationship \\
\bottomrule
\end{tabular}
\begin{tablenotes}
\footnotesize
\item Notes: The dependent variable is the reported size slope. Values in parentheses are \(p\)-values for the coefficient directly above them. Unless otherwise stated, estimates use 1/99 winsorization and country-clustered inference. Positive coefficients imply weaker conventional size premia, negative coefficients stronger premia.
\item The no-winsorization and 5/95 rows report the designated tail-treatment benchmarks. The flat-data row summarizes generalized trust as a persistent country-level environment instead of matching it to the nearest survey year. The precision-weighted limiting diagnostic weights estimates by the inverse of the squared reported standard error. The collapsed row reports ordinary least squares estimates for country-level and study-country-level specifications. The focal coefficient is generalized trust except in the family-trust placebo and residualized-trust rows; in the median-split rows it is the generalized-trust slope above the rule-of-law median. The formal-institution coefficient is rule of law except in the WGI-substitute rows and median-split rows.
\item The reading of the family-trust row comes from the covariance-adjusted marginal effects over the observed rule-of-law support, discussed below.
\item The reduced-control row reports the 0.5 posterior-inclusion-probability threshold specification, which selects standard error, trimming, individual-stock sample, January returns, value controls, momentum controls, price controls, market-characteristic controls, and GDP per capita; the 0.8 threshold drops the market-characteristic control and gives the same qualitative result. Because this reduced specification omits several controls with missing observations, its complete-case sample contains 1,649 estimates from 105 studies and 42 countries.
\end{tablenotes}
\end{threeparttable}
\end{table}

Two rows in Table \ref{tab:robustness} replace the smooth interaction with a median-split low-rule-of-law indicator. In those specifications, the trust coefficient is the trust slope above the rule-of-law median, the low-rule-of-law coefficient is the intercept shift below the median at average trust, and the interaction is the additional trust slope in the lower-rule-of-law regime. Because the dependent variable remains the reported size slope, a positive median-split interaction means that the trust slope is more positive below the rule-of-law median than above it. Whether trust is associated with a smaller premium below the median depends on the sum of the trust and interaction coefficients.

The collapsed specifications answer a different question. They average the estimate-level data to one observation per country, or to one observation per study-country cell, and then re-estimate the institutional interaction on those aggregates. They therefore test whether the same pattern is visible after removing much of the within-country and within-study estimate-level variation that identifies the baseline meta-regression.

Three diagnostics are especially useful for interpreting the interaction. First, when survey-based trust is summarized as a persistent country-level trust environment (the flat-data OLS row), the smooth OLS interaction becomes more negative than in the baseline, \(-0.121\). It is also closer to conventional significance under country clustering, with \(p=0.061\). This is the result one would expect if the relevant trust signal is slow moving. Nearest-survey-year matching is useful for preserving timing, but it can also add measurement noise when trust and legal institutions mainly vary across countries rather than within countries over short horizons. The flat-data check therefore suggests that the pattern is not driven by fine survey timing, and that it is at least as visible when trust is measured as a stable country-level informal institution.

Second, the residualized-trust diagnostic addresses a different concern: that generalized trust is simply proxying for rule of law or broad development. To this end, generalized trust is first purged of its linear association with rule of law, GDP per capita, and GDP growth, and the interaction model is then re-estimated using the remaining component of trust. The interaction again becomes larger in magnitude, \(-0.153\), with \(p=0.051\) under country clustering. The residualized variable cannot be interpreted in the same way as raw generalized trust, so the exercise does not answer the same substantive question as the main specification. Its value is diagnostic: removing the linear component of trust most directly shared with rule of law and development does not eliminate the conditional sign pattern. This makes it less likely that the baseline interaction is only a proxy for broad country development or institutional quality.

Third, the alternative-WGI checks probe the specificity of the formal-institution result. Rule of law was chosen because the literature shows that enforcement, investor protection, legal recourse, and disclosure credibility are the formal mechanisms most relevant for small-firm financing and risk pricing. Regulatory quality, government effectiveness, and control of corruption are broader governance measures with a less direct connection to that mechanism. These substitutes do not reproduce the rule-of-law coefficient. Without a formal comparison across specifications, that pattern is consistent with rule-of-law specificity but also with measurement differences across the correlated WGI indicators.

The baseline pattern is not a by-product of a narrow modeling choice. A reduced-control OLS specification using only the controls with high posterior inclusion probabilities keeps the same signs, although the interaction is weaker than in the full-control specification and it runs on a wider complete-case sample of 1,649 estimates from 42 countries. Removing winsorization leaves the continuous interaction almost unchanged, indicating that the baseline sign pattern is not created by mild tail treatment. The family-trust placebo replaces generalized trust with particularized family trust. Its interpretation comes from the marginal effects and their 90\% confidence intervals. Under study clustering, the family-trust effect is statistically indistinguishable from zero throughout the observed range of rule of law. For generalized trust, the effect is positive and statistically significant where rule of law is low. Under country clustering, family trust is never significantly positive. Its marginal effect becomes significantly negative only within a narrow range at the top of the rule-of-law distribution, which runs against the substitution hypothesis. The family-trust placebo therefore does not show the conditional pattern found for generalized trust.

The interpretation weakens under the remaining checks. The 5/95 winsorized specification materially compresses the support of the dependent variable and attenuates the continuous interaction, making it a deliberately severe stress test. The median-split specifications are secondary: under baseline tail treatment the additional low-rule-of-law trust slope is positive but weak, and under aggressive tail treatment it becomes sharper. This pattern is consistent with a stronger premium-reducing trust association where formal legal quality is weaker, although the threshold evidence depends on both threshold design and tail treatment.

The country-clustered \(p\)-values in Table \ref{tab:robustness} should also be read against the small number and uneven size of country clusters. In the 1/99 baseline, rule of law's country-clustered \(p\)-value rises from 0.016 to 0.058--0.100 under the few-cluster corrections, while trust and the interaction stay well short of significance throughout (Appendix Table \ref{tab:small_cluster_appendix}). Rule of law remains the strongest institutional association but weakens to borderline evidence under the few-cluster checks. The interaction keeps the expected sign without becoming statistically informative.

The precision-weighted limiting diagnostic asks whether the institutional pattern is concentrated among the most precise reported estimates. Re-estimating the baseline by inverse-variance WLS moves the institutional coefficients essentially to zero. We read this as a limiting diagnostic rather than a robustness check, for two reasons. First, inverse-variance weighting is a weak guide to a study's informativeness in observational meta-regression: a reported standard error captures within-study sampling precision. That precision is inflated in large panels with correlated errors and says little about model uncertainty or the between-study heterogeneity that dominates a cross-country sample, so the most heavily weighted estimates need not be the most reliable. Second, the weights here are extremely concentrated: the effective sample size is about 119, and the top 5\% of observations carry about 65\% of total weight. The weighted estimand therefore reflects a small precision-selected subsample rather than the literature as a whole. For these reasons the baseline is deliberately unweighted, and we report the precision-weighted result for completeness rather than as a preferred specification.

Sample composition presents the clearest limitation. There are 49 distinct non-U.S. trust and rule-of-law value pairs in the analysis sample, so the non-U.S. institutional support is thin relative to the estimate count. In the non-U.S. diagnostic, rule of law remains negative and conventionally informative under country clustering, but the continuous interaction falls from about \(-0.108\) to about \(-0.011\) and is far from statistically informative. The rule-of-law association is therefore more stable than the trust-by-rule-of-law interaction. The interaction's generality depends heavily on sample composition. The collapsed country and study-country checks point in the same negative direction as the baseline smooth interaction, but they are statistically weak. The interaction emerges most clearly from the estimate-level meta-regression, which retains variation across study designs and reported estimates; a small cross-country comparison of average slopes supplies little statistical information.

\section{Discussion}\label{sec:discussion}

The results describe institutional heterogeneity in reported size-premium estimates, not the underlying small-minus-big return premium. Rule of law has the more stable institutional association, and its direction runs against the naive prior: stronger rule of law goes with more negative reported size slopes and therefore with larger conventional size premia. The trust result is conditional. Generalized trust is associated with a less negative reported size slope mainly where rule of law is weaker, and that association fades as formal legal quality rises. The OLS interaction is imprecise under clustered inference. The BMA exercise preserves the same posterior sign pattern under control-set uncertainty.

For illustration, we convert the fitted interaction into movements in the reported slope. Using country-level institutional quantiles avoids letting the heavily represented U.S. observations determine the comparison. Appendix Table \ref{tab:institutional_quantiles_appendix} reports the quantiles used for this calculation. Moving generalized trust from the country-level 25th percentile to the 75th percentile is an increase of about 1.45 index units. In a low-rule-of-law environment, around the country-level 25th percentile, the baseline coefficients imply that such a trust increase corresponds to an increase of about 0.18 in the reported size slope. Because more negative slopes imply larger conventional size premia, this is a premium-reducing association. To see how large that movement is, start from the 25th percentile of reported slopes, \(-0.101\) (close to the country-weighted baseline mean). Adding 0.18 gives an implied slope of about 0.08, which sits around the 93rd percentile of the observed slope distribution. The fitted trust movement thus spans the 25th to the 93rd percentile, an interval containing roughly two thirds of the baseline estimation sample.

At the country-level 75th percentile of rule of law, by contrast, the same trust shift implies a fitted movement of about \(-0.04\), reflecting the attenuation and partial reversal that follow from the negative interaction.

The mechanism is most plausible when trust and rule of law are treated as distinct institutional margins rather than as a single scale of ``good institutions.'' The trust literature makes two separate points. At the institutional level, informal constraints help structure exchange alongside formal rules \citep{north1990}, and commercial exchange often requires confidence when contracts cannot specify or enforce every contingency \citep{arrow1972}. At the financial-market level, this logic has concrete margins. Social capital is associated with greater stockholding, check use, and access to institutional credit, with stronger effects where legal enforcement is weaker \citep{guiso2004}. Individual trust predicts stock-market participation through the perceived risk of being cheated \citep{guiso2008}. Societal trust makes corporate earnings announcements more informative to investors \citep{pevzner2015}. These channels make generalized trust a plausible force reducing information and participation frictions around small, opaque firms, though financial-market experience can also feed back into generalized trust itself \citep{jha2025}.

Formal enforcement works through a different margin. Legal protection and enforcement shape investor protection and ownership structures across countries \citep{laporta1998}, and stronger legal institutions and securities regulation are associated with lower implied costs of equity capital \citep{hail2006}. Firm-level evidence also shows why this should matter for size: financial, legal, and corruption obstacles constrain growth most strongly for the smallest firms, and better financial and institutional development weakens those constraints \citep{beck2005b}. But formalization can also be costly in a size-asymmetric way. Evidence from SOX Section 404 suggests that compliance imposed real costs and reduced the market value of small firms \citep{iliev2010}. Bright-line regulatory exemptions gave some small firms incentives to remain below compliance thresholds \citep{gao2009}. The interaction is consistent with a conditional-substitution reading: informal assurance appears most relevant where formal enforcement is weaker. Stronger formal institutions can both relax financing frictions and make small-firm risks or fixed compliance burdens more visible. That does not mean that trust and law are always substitutes. Other settings document complementarity or mixed interactions \citep{bloom2012,carlin2009,cruzgarcia2019,bartling2025}. Our claim is confined to the way reported size-slope estimates line up with institutional environments in this meta-analytic sample.

The diagnostics help interpret the result but also define its scope. The flat-data and residualized-trust checks keep the same interaction sign and produce larger interaction magnitudes, suggesting that the result is not simply an artifact of fine survey timing or of trust proxying mechanically for rule of law and development. The family-trust marginal effects do not reproduce the conditional substitution pattern found for generalized trust. At the same time, the precision-weighted diagnostic is the exception, though a limited one for the reasons discussed above: the pattern is clearest in the unweighted meta-regression and fades under the highly concentrated inverse-variance weights. The non-U.S. and collapsed-sample checks also show that the evidence is too sample-dependent to sustain a broad cross-country claim.

\section{Conclusion}

We ask whether part of the heterogeneity in reported size-premium estimates is organized by institutions. We assemble 1,613 reported size-slope estimates from 105 studies and 31 countries, extending the estimate-level inventory of \citet{astakhov2019}, and match them to generalized trust from EVS/WVS and rule of law from the Worldwide Governance Indicators, with study-design, specification, precision, publication-context, market, and macro-financial controls.

The main institutional result is the rule-of-law association. Over much of the observed trust support, stronger rule of law is associated with more negative reported size slopes and therefore with larger conventional size premia; the association holds under both study- and country-clustered inference and survives dropping the U.S. sample. That direction is consistent with cleaner pricing of small-firm risk where enforcement is strong, and with disclosure and compliance costs that are fixed in size and therefore heavier for small firms. The trust result is conditional: generalized trust is associated with less negative reported size slopes when rule of law is weak. Under the paper's sign convention, that means a smaller conventional size premium where formal institutions are weak. The negative trust-by-rule-of-law interaction indicates that this trust association weakens as rule of law improves.

The evidence offers cautious support for an institutional-substitution interpretation: where formal enforcement is weak, informal assurance can partly stand in for it, and that role recedes as courts and disclosure rules become more reliable. Bayesian model averaging preserves the posterior signs under control-set uncertainty. Among the diagnostics, the rule-of-law association is the more stable of the two. Even so, the rule-of-law association weakens under few-cluster corrections and vanishes under concentrated inverse-variance weighting. That weighting is a weak diagnostic in this observational setting rather than a robustness failure. The trust-by-rule-of-law interaction weakens once the U.S. sample is removed and under precision weighting. Although the analysis neither identifies causal institutional effects nor establishes a general cross-country relationship, trust and rule of law help account for part of the disagreement in reported size-premium estimates. Future work would benefit most from broader non-U.S. evidence and from designs with cleaner identifying variation, such as within-country institutional reforms or cross-listing settings.

\vspace{2\baselineskip}

\noindent\textbf{Data and code availability.} The data and code that reproduce all results in this
paper are available in the replication package at \url{https://meta-analysis.cz/trust} and archived at
\url{https://doi.org/10.5281/zenodo.21486177}.

\vspace{3mm}

\noindent\textbf{Artificial intelligence use.} Claude Opus 4.8 and Claude Fable 5 by Anthropic,
through Claude Code, and GPT-5.6 Sol by OpenAI, through Codex CLI, assisted in preparing and checking
the analysis code, cross-checking and correcting the reported numbers, and editing the text. The data
were collected by hand and without AI. All results were
produced by running the authors' own code on the frozen dataset and can be reproduced from the public
replication package. The authors are responsible for all of the paper's content.

\clearpage
\appendix

\section{Source Inventory and Diagnostic Support}\label{app:inventory}

The source inventory begins with \citet{astakhov2019}'s published-study meta-analysis and its working-paper extension. The main published-study inventory contains 1,746 estimates from 102 studies. Appendix C of that source adds 10 further working-paper studies with 167 additional estimates. The analysis in this paper starts from the source inventory that includes both components, then applies the single-country restriction, the exclusion of estimates with sample midpoints before 1960, the institutional and control-variable matching requirements, and the complete-case restrictions used in the baseline specification. This produces 1,613 estimates from 105 studies and 31 countries. The 105-study count is therefore not a subset of the 102 published studies: all 10 Appendix C working-paper studies remain represented in the final sample, contributing 137 of the included estimates, while some studies from the published inventory do not enter the final complete-case analysis sample. Classifications are frozen at the source inventory's collection date, so an entry listed below as a working paper may since have been published \citep{hou2019}.

\begin{refsection}
\footnotesize
\begin{longtable}{@{}p{0.77\textwidth}r@{}}
\caption{Studies Represented in the Final Analysis Sample, Ranked by Contributed Estimates}
\label{tab:final_study_counts}\\
\toprule
Reference & Included estimates \\
\midrule
\endfirsthead
\caption*{Table \ref*{tab:final_study_counts} continued}\\
\toprule
Reference & Included estimates \\
\midrule
\endhead
\midrule
\multicolumn{2}{r}{Continued on next page} \\
\endfoot
\bottomrule
\multicolumn{2}{p{0.96\textwidth}}{\footnotesize Notes: The table lists the studies represented in the final analysis sample and the number of estimates included from each study. The source inventory follows \citet{astakhov2019}.} \\
\endlastfoot
\fullcite{astakhovstudy041} & 64 \\
\fullcite{astakhovstudy015} & 60 \\
\fullcite{astakhovstudy011} & 54 \\
\fullcite{astakhovstudy038} & 47 \\
\fullcite{astakhovstudy003} & 38 \\
\fullcite{astakhovstudy051} & 38 \\
\fullcite{astakhovstudy034} & 36 \\
\fullcite{astakhovstudy093} & 36 \\
\fullcite{astakhovstudy108} & 36 \\
\fullcite{astakhovstudy014} & 34 \\
\fullcite{astakhovstudy063} & 32 \\
\fullcite{fama1992} & 30 \\
\fullcite{astakhovstudy019} & 28 \\
\fullcite{astakhovstudy064} & 28 \\
\fullcite{astakhovstudy080} & 28 \\
\fullcite{astakhovstudy069} & 26 \\
\fullcite{astakhovstudy084} & 26 \\
\fullcite{astakhovstudy010} & 25 \\
\fullcite{astakhovstudy018} & 25 \\
\fullcite{astakhovstudy025} & 24 \\
\fullcite{astakhovstudy060} & 24 \\
\fullcite{astakhovstudy075} & 24 \\
\fullcite{astakhovstudy081} & 24 \\
\fullcite{astakhovstudy056} & 22 \\
\fullcite{astakhovstudy104} & 22 \\
\fullcite{astakhovstudy107} & 22 \\
\fullcite{astakhovstudy058} & 21 \\
\fullcite{astakhovstudy099} & 21 \\
\fullcite{astakhovstudy013} & 20 \\
\fullcite{knez1997} & 20 \\
\fullcite{astakhovstudy045} & 20 \\
\fullcite{astakhovstudy066} & 20 \\
\fullcite{astakhovstudy072} & 20 \\
\fullcite{astakhovstudy074} & 20 \\
\fullcite{astakhovstudy094} & 20 \\
\fullcite{astakhovstudy095} & 20 \\
\fullcite{astakhovstudy096} & 20 \\
\fullcite{astakhovstudy055} & 19 \\
\fullcite{astakhovstudy097} & 19 \\
\fullcite{astakhovstudy008} & 18 \\
\fullcite{astakhovstudy035} & 18 \\
\fullcite{astakhovstudy085} & 18 \\
\fullcite{astakhovstudy088} & 18 \\
\fullcite{astakhovstudy090} & 17 \\
\fullcite{astakhovstudy022} & 15 \\
\fullcite{astakhovstudy059} & 15 \\
\fullcite{astakhovstudy106} & 15 \\
\fullcite{astakhovstudy087} & 14 \\
\fullcite{astakhovstudy017} & 12 \\
\fullcite{astakhovstudy030} & 12 \\
\fullcite{astakhovstudy082} & 12 \\
\fullcite{astakhovstudy050} & 11 \\
\fullcite{astakhovstudy089} & 11 \\
\fullcite{whited2006} & 10 \\
\fullcite{astakhovstudy031} & 10 \\
\fullcite{astakhovstudy032} & 10 \\
\fullcite{astakhovstudy042} & 10 \\
\fullcite{astakhovstudy048} & 10 \\
\fullcite{astakhovstudy112} & 10 \\
\fullcite{astakhovstudy039} & 9 \\
\fullcite{astakhovstudy057} & 9 \\
\fullcite{astakhovstudy086} & 9 \\
\fullcite{astakhovstudy109} & 9 \\
\fullcite{astakhovstudy111} & 9 \\
\fullcite{astakhovstudy016} & 8 \\
\fullcite{astakhovstudy040} & 8 \\
\fullcite{dichev1998} & 8 \\
\fullcite{astakhovstudy091} & 8 \\
\fullcite{astakhovstudy110} & 8 \\
\fullcite{astakhovstudy020} & 7 \\
\fullcite{astakhovstudy024} & 7 \\
\fullcite{astakhovstudy068} & 7 \\
\fullcite{astakhovstudy028} & 6 \\
\fullcite{astakhovstudy046} & 6 \\
\fullcite{astakhovstudy049} & 6 \\
\fullcite{astakhovstudy062} & 6 \\
\fullcite{astakhovstudy077} & 6 \\
\fullcite{astakhovstudy078} & 6 \\
\fullcite{fama2008} & 5 \\
\fullcite{astakhovstudy027} & 5 \\
\fullcite{astakhovstudy033} & 5 \\
\fullcite{astakhovstudy054} & 5 \\
\fullcite{astakhovstudy092} & 5 \\
\fullcite{amihud2002} & 4 \\
\fullcite{astakhovstudy029} & 4 \\
\fullcite{astakhovstudy043} & 4 \\
\fullcite{astakhovstudy044} & 4 \\
\fullcite{astakhovstudy047} & 4 \\
\fullcite{astakhovstudy052} & 4 \\
\fullcite{astakhovstudy061} & 4 \\
\fullcite{astakhovstudy076} & 4 \\
\fullcite{astakhovstudy101} & 4 \\
\fullcite{astakhovstudy102} & 4 \\
\fullcite{astakhovstudy105} & 4 \\
\fullcite{banz1981} & 3 \\
\fullcite{astakhovstudy065} & 3 \\
\fullcite{astakhovstudy083} & 3 \\
\fullcite{hou2011} & 2 \\
\fullcite{astakhovstudy053} & 2 \\
\fullcite{astakhovstudy073} & 2 \\
\fullcite{astakhovstudy100} & 2 \\
\fullcite{astakhovstudy103} & 2 \\
\fullcite{astakhovstudy113} & 2 \\
\fullcite{astakhovstudy002} & 1 \\
\fullcite{astakhovstudy079} & 1 \\
\end{longtable}
\end{refsection}

\begin{table}[!htbp]
\centering
\caption{Country-Level Institutional Quantiles Used in the Magnitude Calculation}
\label{tab:institutional_quantiles_appendix}
\begin{threeparttable}
\small
\begin{tabular}{lrrr}
\toprule
Variable & P25 & Median & P75 \\
\midrule
Generalized trust & -0.490 & 0.139 & 0.958 \\
Rule of law & 0.171 & 1.056 & 1.625 \\
\bottomrule
\end{tabular}
\begin{tablenotes}
\footnotesize
\item Notes: Quantiles are computed over countries represented in the final analysis sample, rather than over estimate-level observations. They are used only to scale the illustrative fitted magnitudes reported in the text.
\end{tablenotes}
\end{threeparttable}
\end{table}

\begin{table}[!htbp]
\centering
\caption{Small-Cluster Inference for Baseline Institutional Terms}
\label{tab:small_cluster_appendix}
\begin{threeparttable}
\footnotesize
\begin{tabular}{lccc}
\toprule
Term & CR1 \(p\) & CR2/Satt. \(p\) & Wild boot. \(p\) \\
\midrule
Generalized trust & 0.192 & 0.359 & 0.420 \\
Rule of law & 0.016 & 0.100 & 0.058 \\
Generalized trust \(\times\) rule of law & 0.139 & 0.319 & 0.305 \\
\bottomrule
\end{tabular}
\begin{tablenotes}
\footnotesize
\item Notes: The diagnostics use the 1/99 winsorized baseline specification and country clustering. CR1 is the conventional country-clustered reference used in the main table. CR2/Satterthwaite and null-imposed restricted wild-cluster bootstrap \(p\)-values probe the sensitivity of inference to the limited and uneven country-cluster structure. The bootstrap uses Rademacher weights and 99,999 draws.
\end{tablenotes}
\end{threeparttable}
\end{table}

\clearpage

\section{Variable Definitions}

\begin{table}[!h]
\centering
\caption{Variable Dictionary}
\label{tab:variable_dictionary}
\begin{threeparttable}
\footnotesize
\begin{tabular}{p{0.24\textwidth}p{0.68\textwidth}}
\toprule
Variable or group & Definition \\
\midrule
Size slope & Reported coefficient on firm size inherited from \citet{astakhov2019}; more negative values imply a larger conventional size premium. \\
Standard error & Reported or reconstructed standard error of the size-slope estimate. \\
Generalized trust index & Average of standardized EVS/WVS country-year measures of generalized trust. The index combines the standard ``most people can be trusted'' item with survey questions on trust in neighbors, personally known people, people met for the first time, people of another religion, people of another nationality, and people in the respondent's country. \\
Family-trust placebo index & Average of standardized EVS/WVS country-year measures of trust in family, used as a particularized-trust placebo. \\
Rule of law & WGI rule-of-law estimate matched by country and nearest year. \\
Alternative WGI measures & Government effectiveness, regulatory quality, and control of corruption; used in robustness checks. \\
\bottomrule
\end{tabular}
\begin{tablenotes}
\footnotesize
\item Notes: For the generalized- and family-trust indexes, higher values denote higher trust; for rule of law and the alternative WGI measures, higher values denote stronger institutional quality.
\end{tablenotes}
\end{threeparttable}
\end{table}

\begin{table}[!h]
\centering
\caption{Control Variable Groups}
\label{tab:control_dictionary}
\begin{threeparttable}
\footnotesize
\begin{tabular}{p{0.24\textwidth}p{0.68\textwidth}}
\toprule
Group & Variables \\
\midrule
Study design and estimate controls & Relative-size coding, sample midyear, sample length, trimming, OLS and Fama--MacBeth indicators, excess-return coding, individual-stock indicator, January indicator, publication impact, citations, and independent-variable demeaning. \\
Asset-pricing specification controls & Indicators for value, systematic risk, momentum, leverage, profitability, liquidity, volatility, price, market characteristics, market state, information, ownership, growth, distress, and size controls. \\
Macro-financial controls & Bond yield, market-return volatility, private credit to GDP, GDP per capita, and GDP growth. \\
\bottomrule
\end{tabular}
\begin{tablenotes}
\footnotesize
\item Notes: Continuous controls are transformed as described in Section 3.4. Relative to the moderator families named in the main text, the precision moderator (the reported standard error, Table \ref{tab:variable_dictionary}) enters the specification alongside these groups, publication context (publication impact and citations) is listed within the study-design row, and the market-characteristic and market-state indicators within the asset-pricing row.
\end{tablenotes}
\end{threeparttable}
\end{table}

\clearpage
\section{Additional Tables and Figures}

\begin{figure}[!h]
\centering
\IfFileExists{Figures/bma_coefficient_image.pdf}{%
  \includegraphics[width=\textwidth]{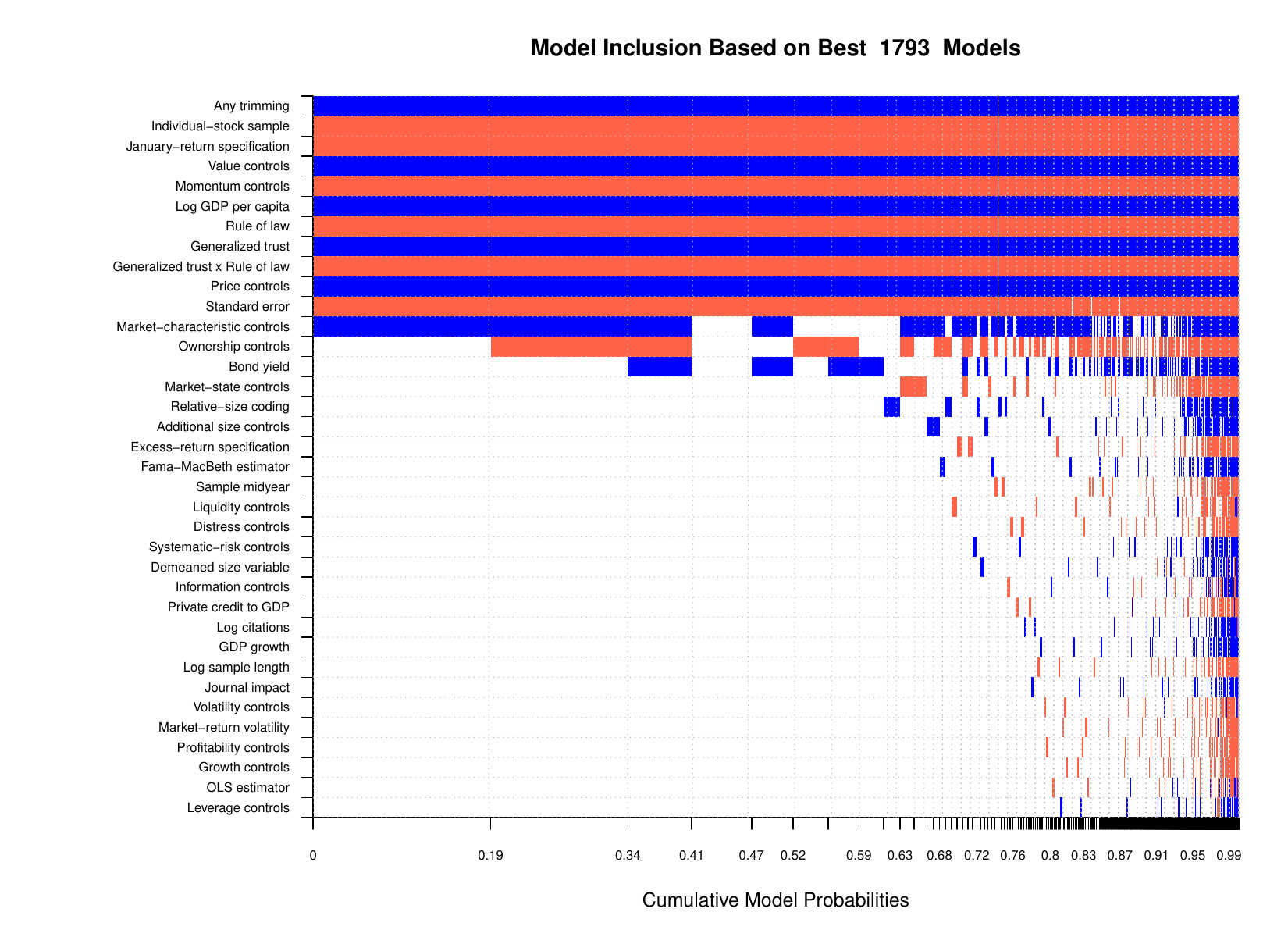}%
}{%
  \fbox{\parbox[c][1.2in][c]{\textwidth}{\centering Placeholder for BMA coefficient image.}}%
}
\caption{Bayesian Model-Averaged Coefficient Summary}
\label{fig:bma_coefficients_appendix}
\par\medskip
\noindent\parbox[t]{\textwidth}{\footnotesize Notes: The figure is the model-inclusion plot of the \texttt{BMS} package. Columns are sampled models ordered by cumulative posterior model probability, rows are variables ordered by posterior inclusion probability, and the plot title reports the number of distinct models the sampler retained. Blue marks inclusion with a positive coefficient, red marks inclusion with a negative coefficient, and white marks exclusion. Institutional terms are forced into every sampled model by design; posterior inclusion probabilities for the remaining controls reflect model uncertainty over the control set.}
\end{figure}

\begingroup
\scriptsize
\setlength{\tabcolsep}{1.6pt}
\begin{longtable}{@{}p{0.31\textwidth}D{.}{.}{-1.3}D{.}{.}{1.3}@{}lD{.}{.}{1.3}@{}lD{.}{.}{-1.3}D{.}{.}{1.3}D{.}{.}{1.3}@{}}
\caption{Full Baseline OLS and BMA Meta-Regression (All Controls)}
\label{tab:full_baseline_ols}\\
\toprule
& \multicolumn{5}{c}{OLS} & \multicolumn{3}{c}{BMA} \\
\cmidrule(lr){2-6}
\cmidrule(lr){7-9}
Variable & \multicolumn{1}{c}{\makebox[0.7in][c]{Coef.}} & \multicolumn{2}{c}{\makebox[0.7in][c]{Study SE}} & \multicolumn{2}{c}{\makebox[0.7in][c]{Country SE}} & \multicolumn{1}{c}{\makebox[0.7in][c]{Post. Mean}} & \multicolumn{1}{c}{\makebox[0.7in][c]{Post. SD}} & \multicolumn{1}{c}{\makebox[0.7in][c]{PIP}} \\
\midrule
\endfirsthead
\caption*{Table \ref*{tab:full_baseline_ols} continued}\\
\toprule
& \multicolumn{5}{c}{OLS} & \multicolumn{3}{c}{BMA} \\
\cmidrule(lr){2-6}
\cmidrule(lr){7-9}
Variable & \multicolumn{1}{c}{\makebox[0.7in][c]{Coef.}} & \multicolumn{2}{c}{\makebox[0.7in][c]{Study SE}} & \multicolumn{2}{c}{\makebox[0.7in][c]{Country SE}} & \multicolumn{1}{c}{\makebox[0.7in][c]{Post. Mean}} & \multicolumn{1}{c}{\makebox[0.7in][c]{Post. SD}} & \multicolumn{1}{c}{\makebox[0.7in][c]{PIP}} \\
\midrule
\endhead
\midrule
\multicolumn{9}{r}{Continued on next page} \\
\endfoot
\bottomrule
\multicolumn{9}{p{0.96\textwidth}}{\scriptsize Notes: The dependent variable is the 1/99 winsorized reported size slope. The OLS coefficient is identical across the OLS columns because both columns estimate the same model; only the cluster-robust covariance matrix changes. Study SE uses study-clustered inference and Country SE uses country-clustered inference. Significance stars correspond to the relevant clustering column: * \(p<0.10\), ** \(p<0.05\), *** \(p<0.01\). BMA columns report posterior means, posterior standard deviations, and posterior inclusion probabilities from the baseline Bayesian model averaging specification. The intercept is always included in BMA, so its PIP is one and no posterior standard deviation is reported by the coefficient summary of the \texttt{BMS} package used for the BMA estimation. The sample contains 1,613 estimates from 105 studies and 31 countries. Several study-design indicators vary mostly within country, so their country-clustered standard errors can be much smaller than their study-clustered counterparts.} \\
\endlastfoot
\addlinespace
Intercept & -1.565 & 0.572 & *** & 0.435 & *** & -1.335 & \multicolumn{1}{c}{--} & 1.000 \\
\addlinespace
\multicolumn{9}{@{}l}{\textit{Precision and coding}} \\
Standard error & -0.139 & 0.306 & & 0.160 & & -0.158 & 0.040 & 0.992 \\
Relative-size coding & 0.051 & 0.048 & & 0.020 & ** & 0.005 & 0.021 & 0.063 \\
\addlinespace
\multicolumn{9}{@{}l}{\textit{Sample period and data treatment}} \\
Sample midyear & -0.002 & 0.002 & & 0.003 & & -0.000 & 0.000 & 0.028 \\
Log sample length & -0.027 & 0.037 & & 0.026 & & -0.000 & 0.002 & 0.013 \\
Any trimming & 0.110 & 0.048 & ** & 0.019 & *** & 0.109 & 0.019 & 1.000 \\
\addlinespace
\multicolumn{9}{@{}l}{\textit{Estimator and return specification}} \\
OLS estimator & 0.001 & 0.033 & & 0.016 & & -0.000 & 0.002 & 0.012 \\
Fama--MacBeth estimator & 0.049 & 0.042 & & 0.021 & ** & 0.001 & 0.009 & 0.036 \\
Excess-return specification & -0.031 & 0.037 & & 0.019 & & -0.001 & 0.005 & 0.027 \\
Individual-stock sample & -0.164 & 0.060 & *** & 0.018 & *** & -0.130 & 0.019 & 1.000 \\
January-return specification & -0.350 & 0.154 & ** & 0.155 & ** & -0.345 & 0.035 & 1.000 \\
\addlinespace
\multicolumn{9}{@{}l}{\textit{Publication context}} \\
Journal impact & 0.001 & 0.037 & & 0.010 & & 0.000 & 0.001 & 0.012 \\
Log citations & -0.005 & 0.015 & & 0.005 & & 0.000 & 0.001 & 0.015 \\
\addlinespace
\multicolumn{9}{@{}l}{\textit{Primary-study controls}} \\
Demeaned size variable & 0.060 & 0.057 & & 0.020 & *** & 0.001 & 0.009 & 0.019 \\
Value controls & 0.150 & 0.047 & *** & 0.029 & *** & 0.139 & 0.020 & 1.000 \\
Systematic-risk controls & 0.012 & 0.031 & & 0.010 & & 0.000 & 0.004 & 0.023 \\
Momentum controls & -0.128 & 0.057 & ** & 0.018 & *** & -0.153 & 0.024 & 1.000 \\
Leverage controls & 0.006 & 0.047 & & 0.031 & & 0.000 & 0.004 & 0.013 \\
Profitability controls & -0.026 & 0.053 & & 0.031 & & -0.000 & 0.004 & 0.015 \\
Liquidity controls & -0.040 & 0.064 & & 0.008 & *** & -0.001 & 0.006 & 0.030 \\
Volatility controls & -0.022 & 0.062 & & 0.014 & & -0.000 & 0.003 & 0.014 \\
Price controls & 0.150 & 0.048 & *** & 0.028 & *** & 0.133 & 0.026 & 0.999 \\
Market-characteristic controls & 0.152 & 0.040 & *** & 0.008 & *** & 0.122 & 0.089 & 0.715 \\
Market-state controls & -0.117 & 0.062 & * & 0.023 & *** & -0.008 & 0.033 & 0.078 \\
Information controls & 0.038 & 0.056 & & 0.026 & & -0.000 & 0.007 & 0.018 \\
Ownership controls & -0.213 & 0.224 & & 0.071 & *** & -0.077 & 0.087 & 0.488 \\
Growth controls & -0.027 & 0.060 & & 0.009 & *** & -0.000 & 0.005 & 0.012 \\
Distress controls & -0.092 & 0.059 & & 0.032 & *** & -0.001 & 0.010 & 0.022 \\
Additional size controls & 0.115 & 0.067 & * & 0.022 & *** & 0.003 & 0.018 & 0.042 \\
\addlinespace
\multicolumn{9}{@{}l}{\textit{Macro-financial controls}} \\
Bond yield & 0.007 & 0.005 & & 0.004 & & 0.002 & 0.004 & 0.290 \\
Market-return volatility & -0.016 & 0.028 & & 0.017 & & -0.000 & 0.002 & 0.013 \\
Private credit to GDP & -0.030 & 0.072 & & 0.069 & & -0.001 & 0.008 & 0.020 \\
Log GDP per capita & 0.175 & 0.069 & ** & 0.050 & *** & 0.140 & 0.026 & 1.000 \\
GDP growth & 0.025 & 0.056 & & 0.063 & & 0.000 & 0.004 & 0.017 \\
\addlinespace
\multicolumn{9}{@{}l}{\textit{Institutional variables}} \\
Generalized trust & 0.146 & 0.086 & * & 0.112 & & 0.135 & 0.066 & 1.000 \\
Rule of law & -0.168 & 0.077 & ** & 0.070 & ** & -0.126 & 0.046 & 1.000 \\
Generalized trust \(\times\) rule of law & -0.108 & 0.073 & & 0.073 & & -0.092 & 0.043 & 1.000 \\
\end{longtable}
\endgroup

\section{Reporting and AI-Use Compliance}
\label{app:reporting}

This appendix records how the paper meets the MAER-Net reporting guidelines for meta-analysis in economics as updated for artificial intelligence \citep{cook2026reporting} and the accompanying principles for AI use \citep{cook2026ai}, and where it departs from them.

\noindent\textbf{Research question and effect size.} Section \ref{sec:background} states the hypotheses and the sign predictions. Section \ref{sec:data} defines the effect size as the reported regression slope on firm size, states the sign convention used throughout, and identifies the reported standard error as the precision measure. Equations \eqref{eq:primary_size_slope} and \eqref{eq:baseline} give the primary-study specification and the meta-regression. The reported slopes are not converted to a common cardinal metric; Section \ref{sec:data} gives the reason, which is that the estimates come from heterogeneous primary specifications and units, so the analysis is carried out, and interpreted, in reported-slope space.

\noindent\textbf{Search, screening, and coding.} We do not conduct a new literature search. The estimate-level inventory, including the search, the inclusion and exclusion rules, and the coding of study and design characteristics, is inherited from \citet{astakhov2019}, which documents how that inventory was assembled; Appendix \ref{app:inventory} states what we take from that source and which restrictions we then apply. This paper therefore supplies no PRISMA flow diagram of its own, and it neither identifies the coders of the underlying estimates nor reports a measure of their agreement. No AI took part in searching, screening, or coding, because none of those steps was carried out here.

\noindent\textbf{Coded variables, meta-regression, and bias.} Table \ref{tab:variable_dictionary} defines the focal variables and Table \ref{tab:control_dictionary} groups the control set; Table \ref{tab:summary_stats} reports descriptive statistics for the outcome, the reported standard error, and the institutional variables, and Table \ref{tab:data_sources} reports sources and coverage. Definitions and descriptive statistics are not tabulated for every coded moderator. The full meta-regression is reported in Appendix Table \ref{tab:full_baseline_ols}, with Bayesian model averaging over the control set as the stated strategy for handling model uncertainty. Publication and precision context enters through the reported standard error and the related moderators; we do not re-estimate publication bias, which \citet{astakhov2019} document for this inventory. Dependence across estimates is handled with study- and country-clustered standard errors and with the small-cluster corrections in Table \ref{tab:small_cluster_appendix}. Precision weighting is probed separately by the inverse-variance diagnostic. The winsorization rule used to limit the influence of outliers is stated in Section \ref{sec:empirical-strategy}.

\noindent\textbf{Interpretation and sharing.} Figures \ref{fig:size_country_boxplot} to \ref{fig:surface} display the distribution of effect sizes and the fitted marginal effects, with the observed support of the conditioning variable shown so that extrapolated regions can be told apart. Table \ref{tab:robustness} reports the robustness checks. Section \ref{sec:discussion} discusses economic significance and the limits of the design. The data and code are archived and citable, and reproduce every reported number.

\noindent\textbf{AI use.} The declaration at the end of the paper names the models, their versions, and the interfaces through which they were used. Data collection was entirely human: the estimate-level inventory was hand-collected in the source study, and the institutional layer was assembled and matched to it by the authors. AI entered only afterwards, and assisted with preparing and checking analysis code, with cross-checking and correcting reported numbers against the output of that code, and with editing prose. It did not select, screen, or code studies; the specifications were chosen by the authors. No table or figure was produced end to end without human validation. Every reported number was produced by running the authors' own code on the frozen dataset. Dates of use and model settings are not recorded for each individual application. The authors are accountable for the content, and can supply prompts and interaction logs to an editor or referee on request.

\noindent\textbf{Departures.} The paper departs from the guidelines at five points. First, reported slopes are not converted to a common cardinal metric. Second, there is no new search, screening, or coding, and so no PRISMA flow diagram, no coder identities, and no inter-coder agreement measure. Third, definitions and descriptive statistics are not tabulated for every coded moderator, and Table \ref{tab:data_sources} does not name provider-level sources for the macro-financial controls. Fourth, there is no new publication-bias investigation; the paper relies on the evidence reported for the source inventory, while still controlling for precision in the meta-regression. Fifth, dates of use and model settings are not recorded for each AI application. The guidelines allow exceptions when accompanied by a rationale. For all but the last departure, the reason is the same: the paper begins from an existing estimate-level inventory, and its contribution is the institutional layer matched to those estimates.

\clearpage

\printbibliography

\end{document}